\documentclass[modern]{aastex631}
\usepackage{amsmath}

\shorttitle{$>$ 1 MeV Flare-Accelerated Electrons}
\shortauthors{Share et al.}
\begin{document}

%\title{Solar Gamma-Ray Observations of an MeV Component of Flare Accelerated Electrons}
% \title {Flare-Accelerated MeV Electrons Revealed by Gamma-Ray Observations} 
%\title {Understanding the Gamma-Ray Continuum from Flare-Accelerated MeV Electrons} 
%\title {{\bf Evidence for a Distinct Coronal Component of Flare-Accelerated MeV Electrons}
\title {Solar Gamma-Ray Evidence for a Distinct Population of $>$ 1 MeV Flare-Accelerated Electrons}

\author[0000-0002-9204-3934]{Gerald H. Share}
\affiliation{ Astronomy Department, University of Maryland, College Park, MD 20740, USA}
\affiliation{ TSC, Resident at the Naval Research Laboratory, Washington, DC 20375-5352, USA}
\author{Ronald J. Murphy}
\affiliation{Retired}
\author{Brian R. Dennis}
\affiliation{Emeritis, Solar Physics Laboratory, Code 671, NASA Goddard Space Flight Center, Greenbelt MD 20771, USA}
\author{Justin D. Finke}
\affiliation{Space Sciences Division, Naval Research Laboratory, Washington, DC 20375-5352, USA}

%\author{Other Authors}
%\affiliation{?, USA}
%\author[0000-0002-0586-193X]{J. Eric Grove}
%\affiliation{Space Science Division, Naval Research Laboratory, Washington, DC 20375-5352, USA}

%% Note that the \and command from previous versions of AASTeX is now
%% depreciated in this version as it is no longer necessary. AASTeX 
%% automatically takes care of all commas and "and"s between authors names.

%% AASTeX 6.31 has the new \collaboration and \nocollaboration commands to
%% provide the collaboration status of a group of authors. These commands 
%% can be used either before or after the list of corresponding authors. The
%% argument for \collaboration is the collaboration identifier. Authors are
%% encouraged to surround collaboration identifiers with ()s. The 
%% \nocollaboration command takes no argument and exists to indicate that
%% the nearby authors are not part of surrounding collaborations.
 
%% Mark off the abstract in the ``abstract'' environment. 
\begin{abstract}

Significant improvements in our understanding of nuclear $\gamma$-ray line production and instrument performance allow us to better characterize the continuum emission from electrons at energies $\gtrsim$ 300 keV during solar flares.  We represent this emission by the sum of a power-law extension of hard X-rays (PL) and a power law times an exponential function (PLexp).  We fit the $\gamma$-ray spectra in 25 large flares observed over 40 years with this continuum and  the calculated spectra of all known nuclear components.  The PLexp is separated spectroscopically from the other components and its presence is required with $>$ 99\% confidence in 18 of the flares.  Its distinct origin is suggested by significant differences between its time histories and those of the PL and nuclear components in 18 of the flares.  {\it RHESSI} imaging/spectroscopy of the 2005 January 20 flare, reveals that the PL and nuclear components come from the footpoints while the PLexp component comes from the corona.  While the index and flux of the anisotropic PL component are dependent on the flare's heliocentric angle, the PLexp parameters do not show comparable dependences with 99.5\% confidence.  The PLexp spectrum is flat at low energies and rolls over at a few MeV.  Such a shape can be produced by thin-target bremsstrahlung from electrons with a spectrum that peaks between 3 -- 5 MeV and by inverse Compton scattering of soft X-rays by 10--20 MeV electrons, or by a combination of the two.  These electrons can produce radiation detectable at other wavelengths.
 
 \end{abstract}
 %This electron population might be responsible for the mysterious submillimeter emission observed in some flares.
 % This summation reproduces the hardening observed in the continuum previously represented by a broken power law but also allows us to determine if the continuum originates in two distinct sources. 

%% Keywords should appear after the \end{abstract} command.
%% The AAS Journals now uses Unified Astronomy Thesaurus concepts:
%% https://astrothesaurus.org
%% You will be asked to selected these concepts during the submission process
%% but this old "keyword" functionality is maintained in case authors want
%% to include these concepts in their preprints.

%% From the front matter, we move on to the body of the paper.
%% Sections are demarcated by \section and \subsection, respectively.
%% Observe the use of the LaTeX \label
%% command after the \subsection to give a symbolic KEY to the
%% subsection for cross-referencing in a \ref command.
%% You can use LaTeX's \ref and \label commands to keep track of
%% cross-references to sections, equations, tables, and figures.
%% That way, if you change the order of any elements, LaTeX will
%% automatically renumber them.
%%
%% We recommend that authors also use the natbib \citep
%% and \citet commands to identify citations.  The citations are
%% tied to the reference list via symbolic KEYs. The KEY corresponds
%% to the KEY in the \bibitem in the reference list below. 

\section{Introduction} \label{sec:intro} 
    
The solar flare $\gamma$-ray spectrum is produced when high-energy electrons and ions interact in the solar atmosphere.  Magnetic reconnection in the corona is widely believed to initiate flares and release the energy to accelerate these particles to relativistic energies in the corona (e.g. \citet{chen20,flei20}) and onto magnetic loops anchored in the photosphere \citep{holm11,benz17}.   Bremsstrahlung from non-thermal electrons in the corona and chromosphere dominates the well-studied X-ray spectrum $\gtrsim$20 keV \citep{denn88,kruc08b,holm11,flet11,whit11,kont11,kont19} and extends into the $\gamma$-ray energy range where nuclear contributions to the spectrum become important \citep{vest88,vilm11}.   It was known from the early $\gamma$-ray measurements in the 1980's by the {\it Solar Maximum Mission} Gamma Ray Spectrometer ({\it SMM}/GRS) that electrons can be accelerated to tens of MeV because their steep $\gamma$-ray spectrum could be distinguished from the hard $\pi$-decay spectrum produced by the interaction of $>$300 MeV protons deep in the chromosphere \citep{forr86}.  
%Since that time $>$50 MeV $\gamma$-ray observations have provided more information on these very-high energy electrons, especially after the {\it Fermi}/LAT instrument was launched \citep{ajel21,shar18}. 
%\citet{suri75} {\bf studied the $\gamma$-ray spectrum of the flare after subtracting background, nuclear lines at 0.5, 1.6, 2.2, 4.4, and 6.1 MeV, and their instrumental continua. They found a residual continuum made up of a power law from 100 keV to 700 keV and an exponential shape $>$ 700 keV.   
 
It has been difficult to unambiguously determine the $>$ 300 keV flare emission produced by electrons, primarily because of the nuclear lines and continuum that also contribute to the $\gamma$-ray spectrum in this energy range.  Nuclear $\gamma$ rays were first detected by the {\it OSO-7} spectrometer from the 1972 August 4 flare \citep{chup73}.  \citet{suri75} found a residual MeV continuum after subtracting background and solar nuclear lines at 0.5, 1.6, 2.2, 4.4, and 6.1 MeV, and their instrumental continua.  However, using the latest nuclear cross sections available at the time, \citet{rama77} concluded that almost all of the residual MeV radiation was due to the superposition of broad and narrow nuclear lines that were not taken into account by \citet{suri75}.  The importance of knowing the nuclear contribution in studies of the MeV electron-produced continuum was highlighted by \citet{vest88} and \citet{shih09}, who found a close correlation of  $>$300 keV electron bremsstrahlung with both nuclear de-excitation and 2.223 neutron-capture line emissions.  Ion acceleration appears to accompany relativistic electron acceleration down to the limiting sensitivity of detectors, thus highlighting the requirement for an accurate knowledge of nuclear-line emission in flares. 

 The road to this understanding commenced with the seminal work of \citet{rama79} and led to the detailed studies of particle acceleration onto magnetic loops \citep{hua89,murp07}, production of the 511-keV positron annihilation line \citep{murp05,murp14}, broad and narrow nuclear de-excitation lines and unresolved continuum \citep{koz02,murp09,murp16,tuns19}, neutrons and the 2.223 MeV capture line \citep{hua87,murp12}, $\pi$-decay $\gamma$-rays \citep{murp87,mack20},  Compton-scattered photons from de-excitation lines and continua \citep{murp18}, and Compton-scattered photons of 2.223 MeV neutron-capture line $\gamma$ rays (Murphy and Share, manuscript in preparation).  The end product of all of this work is an array of nuclear $\gamma$-ray spectral templates for each of the processes for different ambient and accelerated particle abundances, flare locations, and ion spectral indices.  Access to these templates is provided by OSPEX (Object Spectral Executive)\footnote{\url{https://hesperia.gsfc.nasa.gov/ssw/packages/spex/doc/ospex_explanation.htm}} that is available in the SSW IDL software depository\footnote{\url{http://www.lmsal.com/solarsoft/ssw_whatitis.html}.}

%, with almost no contribution from other mechanisms.  It is ironic that 50 years later, with improved instrumentation and understanding of the detectors, comprehensive knowledge of nuclear $\gamma$-ray production, and sophisticated analysis techniques, that we confidently report the discovery of the similarly-shaped distribution of electrons suggested by \citet{suri75}.

%The presence of nuclear lines continues to compromise studies of the MeV electron continuum. \citet{kong13} studied power-law spectra of flares in the GRS catalog \citep{vest99} that break-up near 1 MeV but have no evidence for nuclear lines.  However, we show in $\S$\ref{subsec:weak} that the summed spectrum from many of the events in their study has a significant nuclear-line contribution $>$1 MeV, compromising their conclusions.   \citet{dasi21} fitted both {\it RHESSI} and {\it Fermi}/GBM spectra near the end of the impulsive phase of the 2013 May 13 flare and found that the index obtained from the radio spectrum agrees with the X-ray derived index for energies above a break energy of $\sim$600 keV.  \citet{shar18} studied this flare in their catalog of Long Duration Gamma Ray Emission events $>$100 MeV.  They noted that the {\it RHESSI} spectra show evidence for nuclear-line emission and that the GBM spectral data $\gtrsim1$ MeV were contaminated by radioactivity produced by protons in a recent passage through the South Atlantic Anomaly.  

Using this newly-derived information on nuclear $\gamma$-ray production, we began a program to fit the time-integrated spectra from 25 intense nuclear-line flares to determine the ambient and accelerated elemental abundances in the plasma where protons and heavier ions interact. It soon became clear that the broken power-law function commonly used in previous studies to represent the electron-generated $\gamma$-ray continuum was inadequate to fit all of the   flare spectra.  This was in part due to the fact the continuum rolled over at high energies.  We therefore replaced the broken power-law with the sum of two continuum components: a Power Law extension of the hard X-rays to MeV energies (PL) and a hard MeV continuum represented by a flat Power Law times an exponential (PLexp).  This more generalized form naturally accommodates spectra that harden in the MeV range, such as power laws that break up, and importantly allows for possibility that it is produced by two different sources, such as the spectrally distinct footpoint and coronal emissions observed in the 2005 January 20 flare \citep{kruc08}.  The exponential factor accommodates spectra with rollovers at high energy \citep{elli85}.   This summed continuum was first used in spectral fits to the 2010 June 12 flare observed by {\it Fermi}/GBM \citep{acke12a} and in more recent studies \citep{kurt17,murp18,lyse19}.

In this paper we discuss the characteristics of the PLexp continuum in solar flares over three decades of intensity.  We present compelling evidence that it is a unique component, distinct from both the power-law extension of the hard X-rays and the nuclear $\gamma$-rays, and that it reveals a new population of MeV electrons.  In $\S$ref{sec:fits} we describe the instruments used and our fits to the spectra in 25 $\gamma$-ray line flares. In $\S$\ref{sec:distinct} we detail the spectral, temporal, spatial, and directional studies that reveal the distinct origin of the PLexp emission. We discuss how the spectral characteristics of the PLexp component in weak flares and `electron-dominated' episodes \citep{rieg98} differ from those in 25 strong nuclear-line flares in $\S$\ref{sec:plexp_par}.  We estimate the spectral characteristics of this new population of MeV electrons in $\S$\ref{sec:origin} assuming that the PLexp component is produced by thin-target bremsstrahlung and/or inverse Compton scattering in the corona.   In $\S$\ref{sec:disc} we summarize the evidence for these coronal MeV electrons and their possible origins, and suggest that they may explain some puzzling solar observations.

\section{Fits to the $\gamma$-ray Spectra in Large Nuclear-Line Flares}\label{sec:fits}

We use data from four instruments: the {\it Solar Maximum Mission (SMM)} Hard X-ray Burst Spectrometer (HXRBS) \citep{orwi80} and Gamma Ray Spectrometer (GRS) { \citep{forr80},  the {\it Ramaty High Energy SpectroScopic Imager (RHESSI)} \citep{lin02}, and the {\it Fermi} Gamma Ray Burst Monitor (GBM) \citep{meeg09}.  The only GRS data that are preserved for use with modern computers are limited to flares observed $>$300 keV, and the associated backgrounds. Unfortunately, in some cases a portion of the flare or background is missing.  Therefore not all of the flares can be studied in their entirety.  All of the data are in a format compatible with SSW IDL software and were analyzed with OSPEX.  We have identified issues with how the original {\it SMM} GRS and {\it RHESSI} detector response  matrices (DRMs) accounted for $\gamma$ rays that do not lose all of their energy in the instrument.  This affects some of the nuclear-line studies reported up to the time of the solar $\gamma$-ray review by \citet{vilm11}.  The DRMs of both instruments were corrected and incorporated into OSPEX shortly thereafter.  We detail the changes made to the GRS DRM in Appendix \ref{sec:resp}.   A normalization error in the Compton component of direct photon interactions accounted for half of the problem in the {\it RHESSI} DRM, with a coincidence mode issue accounting for the other half (David Smith 2009, private communication).

As we discussed above, we fit $>$300 keV flare spectra with the sum of a nuclear component and a non-nuclear continuum. The non-nuclear continuum is represented by the sum of two components, 1) a Power Law (PL) function and 2)  a Power Law times an exponential (PLexp)}, rather than the broken power-law function used in previous studies.   It has the following form for the photon flux, N, as a function of photon energy, E

\begin{equation}
dN/dE = A_{PL}(E/E_0)^{-S_{PL}} + A_{PLexp}(E/E_0)^{-S_{PLexp}} exp^{-E/E_R}  
\end{equation}

where A$_{PL}$ and S$_{PL}$ are the amplitude and index of the PL component, A$_{PLexp}$ and S$_{PLexp}$ are the amplitude and index of the PLexp component, E$_0$ is the normalization energy, (we used 0.3 MeV for the fits in this paper), and E$_R$ is the exponential rollover energy.  

For the nuclear component, we use the nuclear $\gamma$-ray spectra \citep{murp09} calculated for the best-fitting elemental and accelerated abundances and ion spectral indices derived from our nuclear studies.   Our fits to all the GRS flare spectra and to those {\it RHESSI} spectra without significant degradation due to radiation damage included the following nuclear components: $\alpha$-$^4$He fusion lines, narrow, broad, and $^3$He-induced de-excitation line templates, a 2.223-MeV Gaussian neutron-capture line and templates representing its solar Compton-scattered continuum, and the 511-keV positron annihilation line and positronium continuum.  Due to the poorer spectral resolution of the {\it Fermi}/GBM instrument we included only the 511-keV line, the 2.223-MeV line and its solar scattered component, and narrow- and broad-line templates.   The fits also included $\pi$-decay emission extending down below 10 MeV in the {\it RHESSI} and {\it Fermi}/GBM spectra and in one flare observed by GRS.  This emission comes from interactions of $>$300 MeV protons accelerated in the impulsive and late phases of flares \citep{shar18,ajel21}.  Because GRS spectroscopy only extends to 8.5 MeV, it is difficult to detect and fit the $\pi$-decay component.  We only included a $\pi$-decay component in our fits to SOL19891019T12:59 where a significant flux of $>$10 MeV late-phase emission was observed throughout the observation \citep{shar22}.

\begin{figure}[h!]
\gridline{\fig{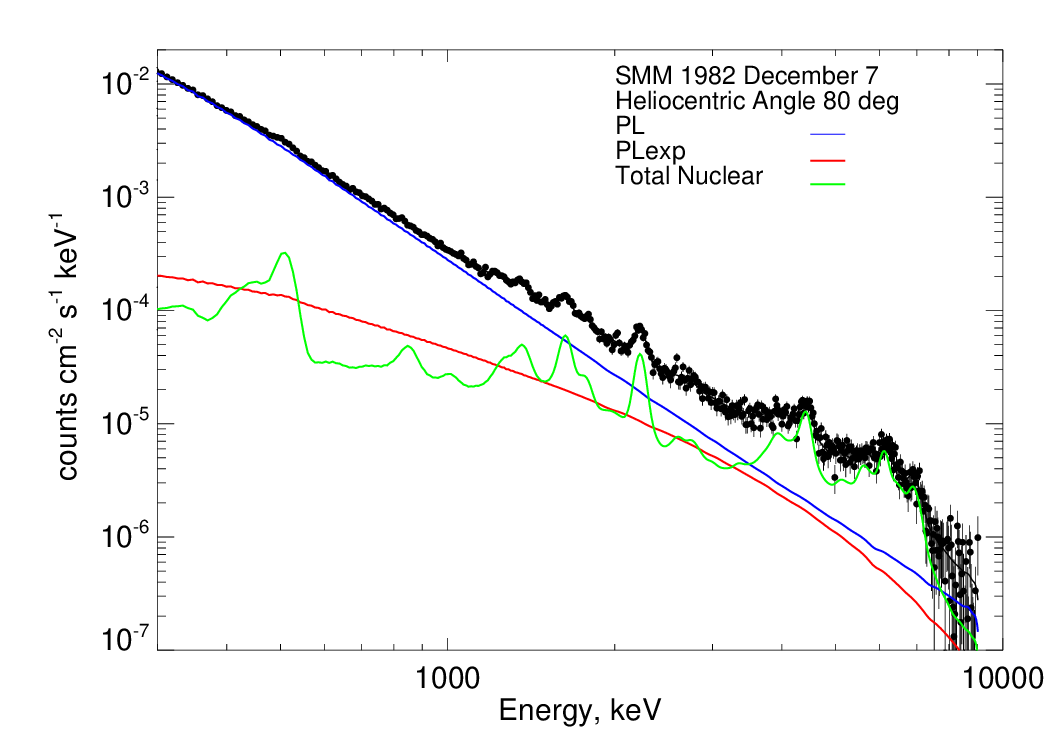}{0.5\textwidth}{(a)}
         \fig{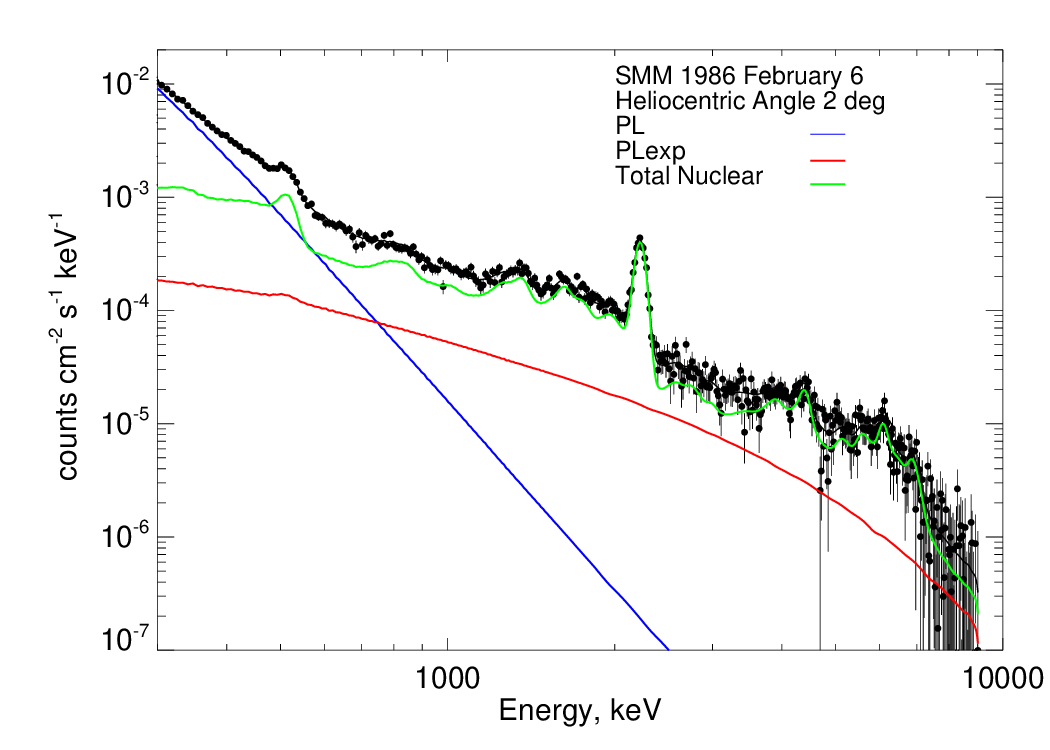}{0.5\textwidth}{(b)}}
\gridline{\fig{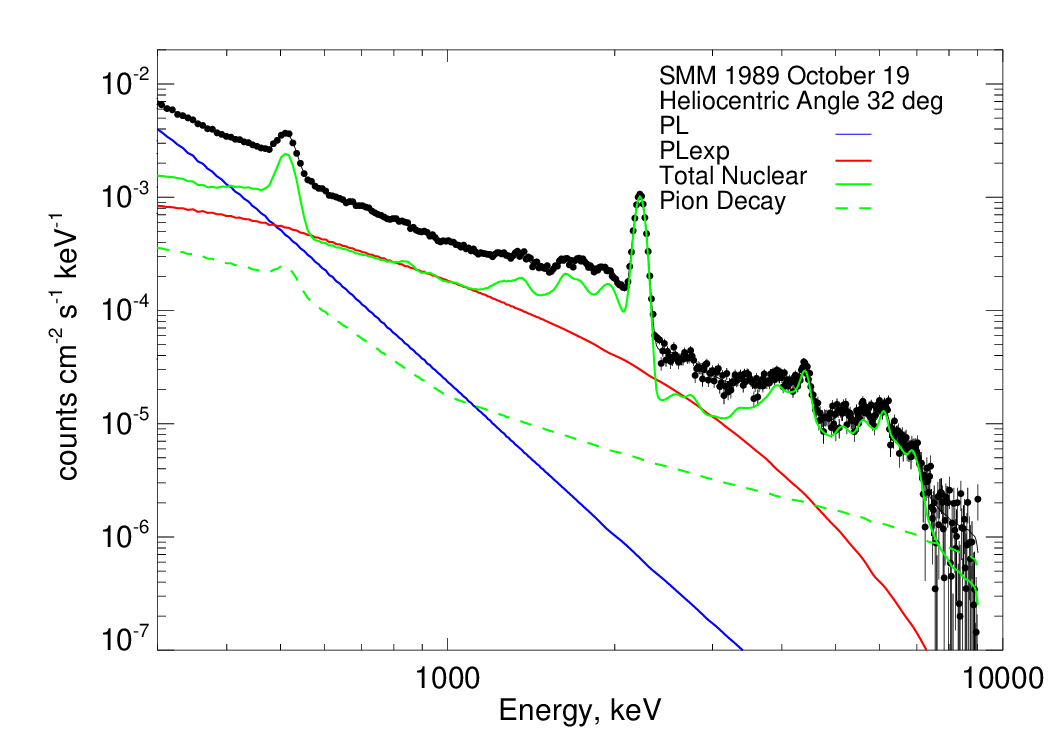}{0.5\textwidth}{(c)}
         \fig{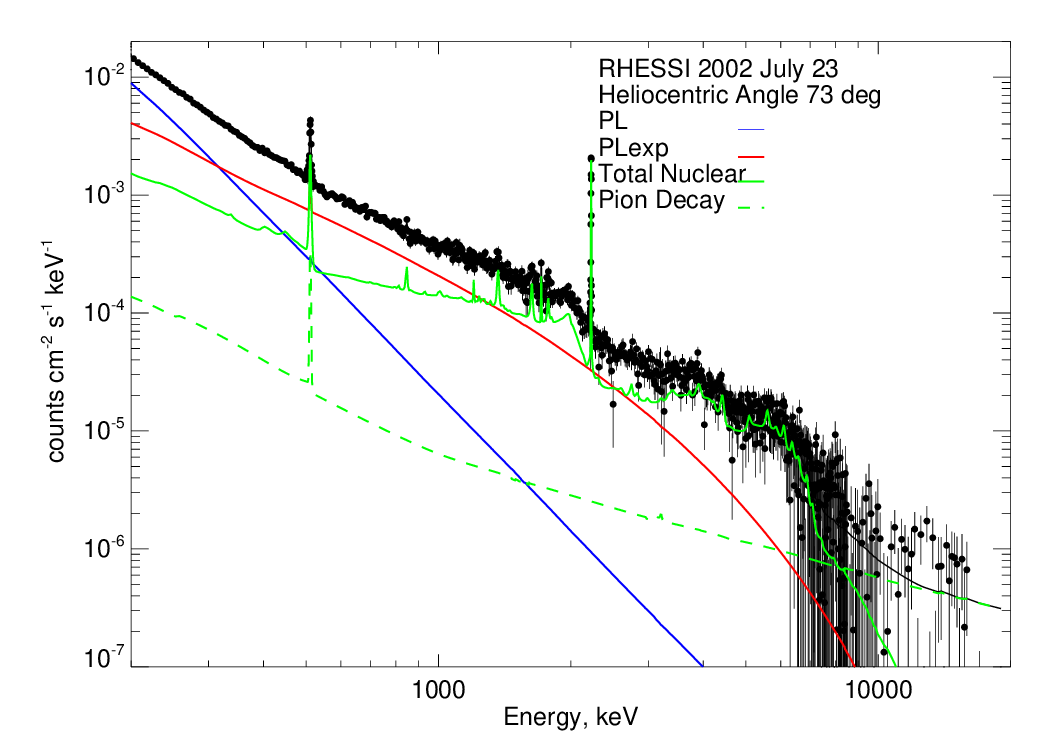}{0.5\textwidth}{(d)}}        
\caption{Counts spectra of four flares well fit by Power Law (PL, blue), Power Law times exponential (PLexp, red), and total nuclear (green) components. Also plotted in panels  (c) and (d) are the $\pi$-decay (dashed green curves) fits to the 1989 October 19 spectrum observed by GRS and the 2002 July 23  spectrum observed by {\it RHESSI}. 
\label{ctsspectra}}
\end{figure}

\begin{figure}[h!]
\gridline{\fig{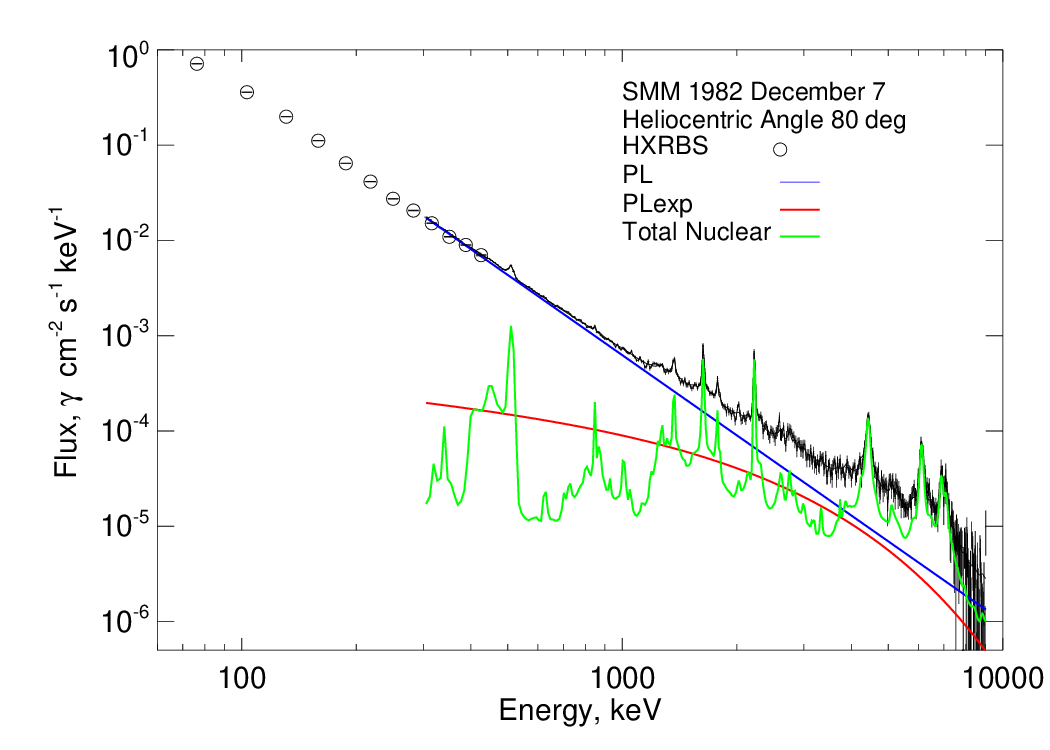}{0.5\textwidth}{(a)}
         \fig{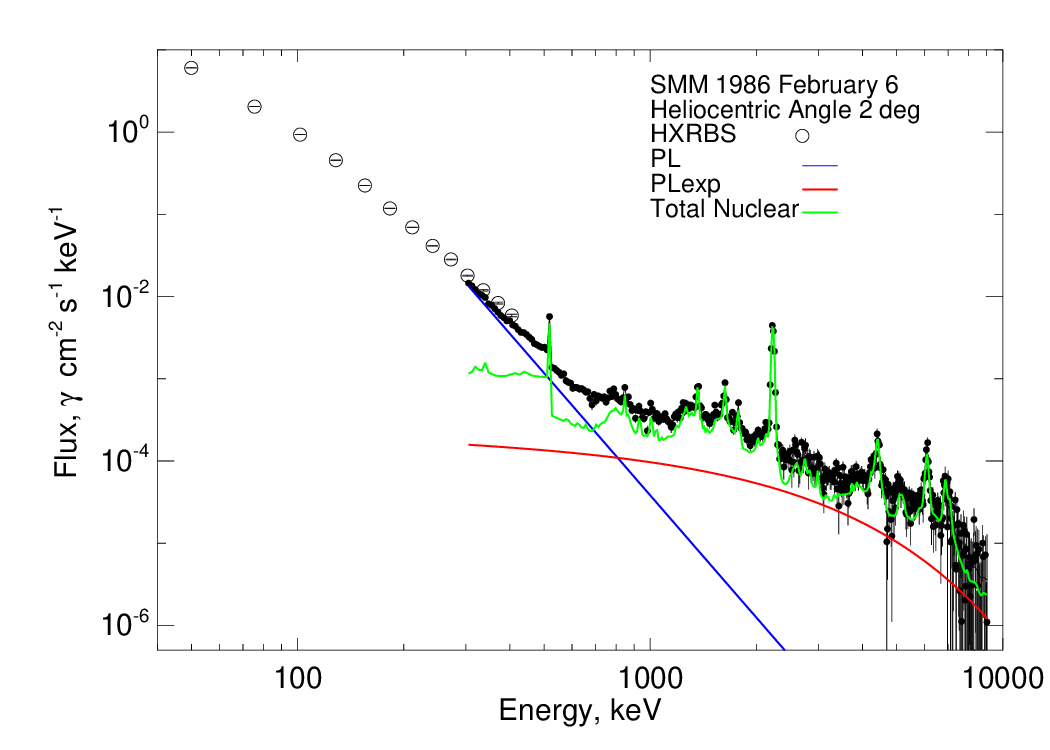}{0.5\textwidth}{(b)}}
\gridline{\fig{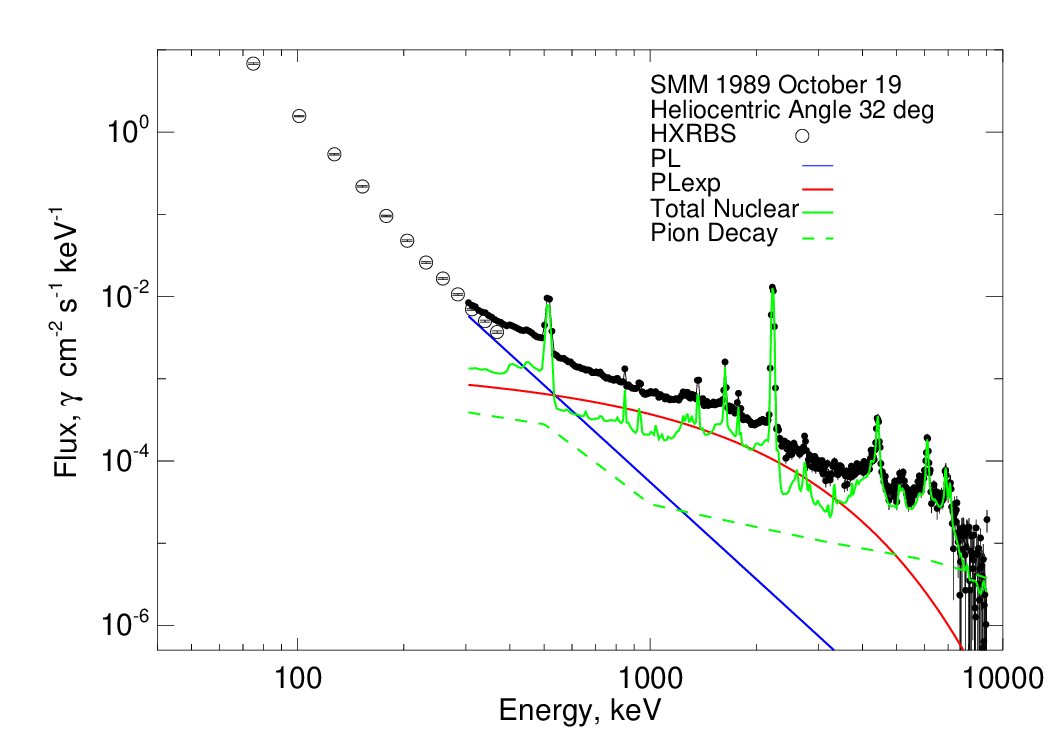}{0.5\textwidth}{(c)}}
%         \fig{02jul23_rhessi.eps}{0.5\textwidth}{(d)}}        
\caption{Photon spectra of three flares covering the range from 60 keV to 8.5 MeV using combined data from {\it SMM} GRS and HXRBS.  They are well fit above 300 keV by PL (blue), PLexp (red), and total nuclear (green) components.  Also plotted in panel  (c) is the $\pi$-decay spectrum (dashed green curve) fit to the 1989 October 19 observed by GRS.  The plots demonstrates that the PL  component observed $>$300 keV by GRS is the extension of hard X-ray bremsstrahlung. We note that the significance of narrow-lines is exaggerated in plots of photon spectra due to their compliance with the fitted model \citep{feni83}.
\label{phspectra}}
\end{figure}

In Figure \ref{ctsspectra} we plot our fits to the time-integrated count spectra of three large flares observed by GRS and one by {\it RHESSI}.  Counts spectra include instrumental contributions such as escape peaks and continua from partial energy losses.  The green curve shows the total nuclear contribution, including the narrow, broad, and unresolved nuclear de-excitation lines, the 511 keV annihilation line, and the 2.223 MeV neutron-capture line and its scattered solar component.  The PL (blue) and PLexp (red) components are clearly distinguished from each other and from the nuclear component.  In panels (c) and (d) we also show the fitted $\pi$-decay components (dashed green curves), which have a flat spectrum that is distinguishable from the MeV rollover in the PLexp component.   

In establishing the distinct nature of the PLexp, it is important to demonstrate that the PL function is, in fact, the extension of the hard X-ray emission observed below 300 keV.  For that reason, in Figure \ref{phspectra} we combine our derived {\it SMM} HXRBS and GRS photon spectra to cover the energy range from 60 keV to 8500 keV in three flares.  In obtaining the HXRBS photon spectra we fit the data with a broken power-law spectral shape. The plots show the excellent agreement between the two instruments in their overlapping energy range and confirm that the PL component derived in GRS fits $>$ 300 keV is indeed the extension of the hard X-ray spectrum. The plotted components are from our fits to the spectra $>$ 300 keV.   We note that there is evidence in some HXRBS spectra for deviations from a single hard X-ray power law that requires further study.  

The PL component plotted in Figures \ref{ctsspectra} and \ref{phspectra} exhibits flare-to-flare variability, displaying striking changes in its hardness and intensity, relative to the PLexp and nuclear components.  As discussed in $\S$\ref{subsec:iso}, we attribute this behavior to anisotropy of the electrons producing the PL component.  In contrast, the PLexp component is relatively constant in shape and intensity when compared to the nuclear component.  These different behaviors suggest that the PL and PLexp components have different origins.
%Had we fit the spectra with broken power laws, the break energies would have been $\sim$ 400 keV in two flares, $\sim$ 600 keV, and $\sim$ 3 MeV. 

%\begin{rotatetable}
%\movetableright=-3cm
\begin{deluxetable*}{ccccccccccc}[h]
%\begin{deluxetable*}{ccccccccccc}
\tablenum{1}
\tablecaption{Details of the Fits to 25 Nuclear-Line Flares \label{tab:sum}}
\tablewidth{0pt}
\tablehead{
\multicolumn{3}{c}{Flare} &\multicolumn{3}{c}{Power Law}  &\multicolumn{4}{c}{Power Law $\times$ Exp} & Nuclear  \\
\colhead{Date/Time}  & \colhead{Dur} & \colhead{Angle} & \colhead{Amp}  &  \colhead{Index} &  \colhead{Flux}& \colhead{Amp}  & \colhead{Index} & \colhead{Energy } & \colhead{Flux } & \colhead{Flux} \\
\colhead{SOL}  & \colhead{s} & \colhead{deg} & \colhead{A$_{PL}$\tablenotemark{a} }  &  \colhead{S$_{PL}$} &  \colhead{Flux$_{PL}$\tablenotemark{b}}& \colhead{A$_{PLexp}$\tablenotemark{a} }  & \colhead{S$_{PLexp}$} & \colhead{E$_R$\tablenotemark{c} } & \colhead{Flux$_{PLexp}$\tablenotemark{b} } & \colhead{Flux$_{nuc}$\tablenotemark{b} } \\ 
\colhead{(1)}  & \colhead{(2)} & \colhead{(3)} & \colhead{(4)}  &  \colhead{(5)} &  \colhead{(6)}& \colhead{(7)}  & \colhead{(8)} & \colhead{(9) } & \colhead{(10) } & \colhead{(11) }  
}
%\colnumbers
\startdata
19810410T16:45 &  573 &    38 &    2.90(0.18)&   3.4(0.2)&   0.36(0.04)&   0.06(0.17)&   0.4(1.7) &  4.3(6.3)&   0.09(0.25)&   0.10(0.02)\\
19810427T07:56 &  2425 &    91 &    2.73(0.19)&   3.4(0.2)&   0.34(0.04)&   0.37(0.19)&   0.9(0.3) &  2.7(0.8)&   0.22(0.11)&   0.19(0.02)\\
%19820603T11:45&  1195 &    72 &    1.39(0.55)&   4.60(1.88)&   0.12(0.08)&   0.57(0.48)&   0.84 &  1.55( 1.64)&   0.25(0.21)&   0.07(0.02)\\
19820709T07:35&  197 &    73 &   32.40(1.85)&   3.2(0.2)&   4.52(0.42)&   3.46(1.90)&   0.8(0.4) &   2.0(0.6)&   1.97(1.08)&   0.51(0.07)\\
19821126T02:31&  262 &    87 &   10.94(1.44)&   3.6(0.4)&   1.27(0.25)&   1.92(1.51)&   0.7(0.7) &  1.3(0.7)&   0.81(0.64)&   0.27(0.03)\\
19821207T23:39&  2080 &    80 &   18.11(0.15)&   2.8(0.1)&   3.02(0.06)&   0.23(0.18)&   0.3(0.4) &  1.8(0.5)&   0.21(0.16)&   0.20(0.01)\\
%19840425T00:03& 852 &    45 &    2.78(0.20)&   5.83(0.78)&   0.17(0.03)&   0.40(0.14)&   0.10 &  1.44( 0.49)&   0.40(0.14)&   0.28(0.03)\\
19860206T06:19 & 377 &     2 &   14.62(0.29)&   5.0(0.2)&   1.11(0.05)&   0.18(0.14)&   0.1(0.5) &  1.9(0.5)&   0.26(0.20)&   0.56(0.05)\\
19881216T08:42\tablenotemark{e} & 2766 &    43 &    3.30(0.10)&   3.4(0.1)&   0.40(0.03)&   0.16(0.13)&   0.5(0.5) &  2.9(0.9)&   0.15(0.13)&   0.15(0.01)\\
19890306T14:14\tablenotemark{e}&  2351 &    76 &   18.54(0.13)&   2.9(0.1)&   2.98(0.03)&   0.13(0.06)&   0.1(0.3) &  1.9(0.3)&   0.17(0.08)&   0.32(0.01)\\
19890310T19:03& 2767 &    44 &    5.16(0.18)&   3.4(0.1)&   0.66(0.04)&   0.34(0.22)&   0.8(0.4) &  2.4(0.6)&   0.22(0.14)&   0.14(0.01)\\
19890317T17:34&   410 &    68 &   18.25(1.62)&   3.4(0.2)&   2.25(0.28)&   4.53(1.64)&   1.3(0.3) &  3.5(0.7)&   1.89(0.69)&   0.62(0.07)\\
19890503T03:51& 590 &    44 &    2.87(0.75)&   4.1(0.9)&   0.28(0.11)&   0.77(0.79)&   1.5(0.7) &  3.8(3.2)&   0.27(0.28)&   0.13(0.02)\\
19890816T01:22&803 &    87 &    3.19(0.38)&   2.4(0.3)&   0.68(0.15)&   0.28(0.42)&   0.3(0.9) &  1.7(1.1)&   0.28(0.41)&   0.17(0.02)\\
19890817T00:47& 2280 &    90 &    6.88(0.33)&   2.3(0.1)&   1.58(0.10)&   1.08(0.32)&   0.5(0.2) &  1.5(0.3)&   0.68(0.20)&   0.08(0.01)\\
19890909T09:09 &  311 &    30 &   10.93(0.23)&   3.9(0.1)&   1.14(0.06)&   0.39(0.31)&   0.7(0.5) &  3.7(1.4)&   0.37(0.29)&   0.22(0.03)\\
19891019T12:59&  1170 &    32 &    6.19(0.38)&   3.9(0.3)&   0.64(0.07)&   1.13(0.28)&   0.1(0.3) &  1.1(0.2)&   0.76(0.19)&   0.46(0.02)\\
19891024T17:57&  868 &    64 &   13.48(1.09)&   3.5(0.2)&   1.60(0.20)&   1.10(1.04)&   0.7(0.8) &  1.4(0.8)&   0.49(0.47)&   0.19(0.02)\\
19891115T19:31 & 836 &    37 &    6.26(0.26)&   3.6(0.2)&   0.73(0.06)&   0.22(0.25)&   0.6(0.7) &  2.2(1.1)&   0.16(0.18)&   0.20(0.03)\\
20020723T00:27& 960 &    73(35)\tablenotemark{d}  & 13.17(3.00)&  3.7(0.7)&  1.47(0.53)&   4.01(3.33)&   0.9(0.8) &   1.5(0.9)&   1.56(1.30)&   0.49(0.05)\\
20031028T11:08&  480 &    18 &   30.26(1.54)&   4.9(0.5)&   2.36(0.32)&  12.15(1.51)&   0.9(0.1) &   2.2(0.2)&   6.45(0.80)&   1.37(0.12)\\
20031102T17:16&  460 &    53 &   52.81(1.23)&   3.4(0.1)&   6.52(0.39)&   1.25(1.33)&   0.3(0.6) &   1.93(0.6)&   1.24(1.32)&   1.64(0.16)\\
20050120T06:44&  1080 &    62 &   25.70(2.97)&   3.3(0.3)&   3.59(0.69)&   6.90(3.54)&   1.0(0.3) &   2.8(0.5)&   3.38(1.73)&   0.61(0.08)\\
20061206T18:42&  1084 &    60 &   27.23(2.49)&   3.2(0.3)&   3.76(0.67)&   0.91(1.52)&   0.5(0.9) &   2.5(1.0)&   0.87(1.46)&   0.49(0.09)\\
20100612T00:55&     50 &    61 &   48.66(1.11)&   3.3(0.1)&   6.23(0.27)&   1.56(1.24)&   0.8(0.5) &   2.4(0.9)&   0.97(0.77)&   0.44(0.05)\\
%20140225T00:43&  489 &    78 &   65.60(1.55)&   3.0(0.1)&  10.01(0.35)&   4.94(1.78)&   1.0(0.2) &   2.4(0.4)&   2.33(0.84)&   0.43(0.02)\\
20140225T00:43&   489 &    78 &   56.75(4.46)&   3.2(0.1)&   7.72(0.72)&  14.67(4.74)&  1.6(0.2) &  3.8(0.7)&   4.62(1.49)&   0.42(0.03)\\
20170910T15:53& 1034 &    91 &    5.98(0.57)&   2.5(0.1)&   1.17(0.15)&   0.65(0.59)&   0.4(0.8) &   1.2(0.6)&   0.37(0.34)&   0.11(0.02)\\
   \\
\enddata
\tablecomments{See Equation 1 for definition of the parameters but note that our energy units here are in MeV for space considerations}
\tablenotetext{a}{$\gamma$ cm$^{-2}$ s$^{-1}$ MeV$^{-1}$}
\tablenotetext{b}{$\gamma$ cm$^{-2}$ s$^{-1}$}
\tablenotetext{c}{Rollover Energy, MeV}
\tablenotetext{d}{Field lines appear to be tilted from the vertical by $\sim$40$^{\circ}$ \citep{smit03}}
\tablenotemark{e}{Spectral accumulation was delayed until after the time when $\pi$-decay emission was detected by the GRS high-energy matrix}
\end{deluxetable*}
%\end{rotatetable}
%
As the OSPEX software has been used to fit RHESSI data in hundreds of publications, we use it in our study.  The results of the OSPEX  fits to 25 time-integrated flare spectra are listed in Table \ref{tab:sum}\footnote{The flares observed by {\it SMM} on 1982 June 3 and 1984 April 24/25 had significant dead time during the impulsive phase and consequently were not included in the table and ensuing analyses.}.   We list the following parameters in the numbered columns: (1) the IAU identifier for the flare \citep{leib10}, where the time is the start time of the accumulation; (2) the duration of the observation;  (3) the heliocentric angle of the flare from Sun center; (4)--(6) the parameters (uncertainties) of the fitted PL, amplitude A$_{PL}$, spectral index S$_{PL}$, and integrated flux from 0.3 to 10 MeV, Flux$_{PL}$;  (7)--(10) the parameters (uncertainties) of the fitted PLexp, amplitude A$_{PLexp}$, spectral index S$_{PLexp}$, exponential rollover energy E$_R$, and integrated flux from 0.3 to 10 MeV, Flux$_{PLexp}$, and (11) the total narrow plus broad nuclear de-excitation line flux (uncertainties), Flux$_{nuc}$.  

The uncertainties of the PLexp amplitude, index, and rollover energy are large because all were free parameters in the fits and they are strongly correlated with one another and with the PL parameters.  For example in our fits to the 2003 October 28 flare observed by {\it RHESSI} the OSPEX software reported that there is a 90\% correlation between the PLexp amplitude and index and there is an 84\% correlation between the PL index and both the PLexp amplitude and index. The correlation between the PL amplitude and the PLexp parameters is $\sim$ 40\%.  Although we find that these strong correlations are reflected in the reported OSPEX uncertainties, there are questions about how accurately they are calculated (Ireland et al. 2013). 

Even though there are large uncertainties in the three parameters, we found that our fits require the presence of the PLexp component, or one with a similar shape, with high probability in most flares.  We did this by using the $\chi$$^2$ statistic to measure the quality of the fits to the spectral data.  No systematic  uncertainties were added to the data points when doing the fits.  We then compared the values of $\chi$$^2$ for the fits used to obtain the parameters in Table \ref{tab:sum} with the values obtained in fits to the flare spectra with no PLexp component.  For example, in our fits to the 1988 December 16 flare we obtained a $\chi$$^2$ of 446 for 413 degrees of freedom (dof) with the PLexp component and a $\chi$$^2$ of 547 for 416 dof without it.  The fit had a probability of 13\% with the PLexp component and 1.5 $\times$ 10$^{-3}$\% without it.  This indicates that a component with the characteristics of the PLexp component is required with a confidence $>$99.99\% for this flare.  In a similar study of all 25 flares we found that the PLexp component is required with $>$ 99\% confidence in 18 of the flares.  

\section{Evidence That the PLexp Continuum is a Distinct Component}\label{sec:distinct}

In the three sections below we present evidence that the PLexp component has a different origin from the power-law extension of the hard X-ray emission (PL).

%\subsection{Flare-to-Flare Spectral Variation}\label{subsec:flr2flr}

%Discuss how the flare to flare spectral variation of the PL, Plexp, and nuclear point to its distinct nature

\subsection{Temporal Variation in Flares}\label{subsec:temp}
\begin{figure}[h!]
\gridline{\fig{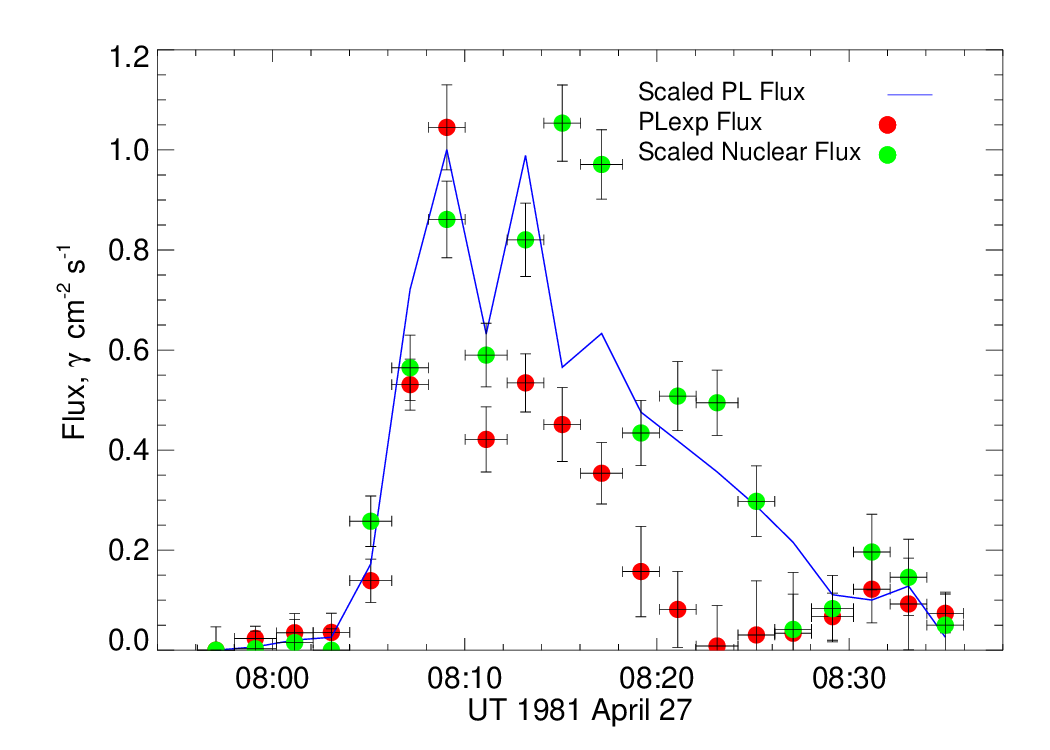}{0.5\textwidth}{(a)}
         \fig{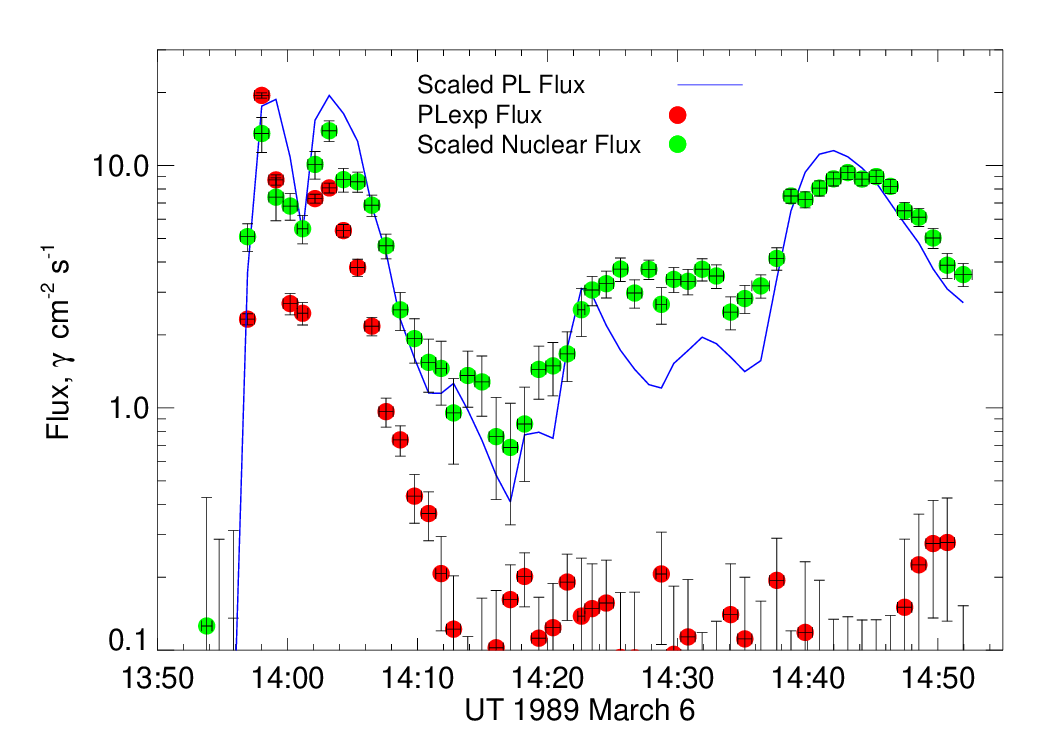}{0.5\textwidth}{(b)}}
\gridline{\fig{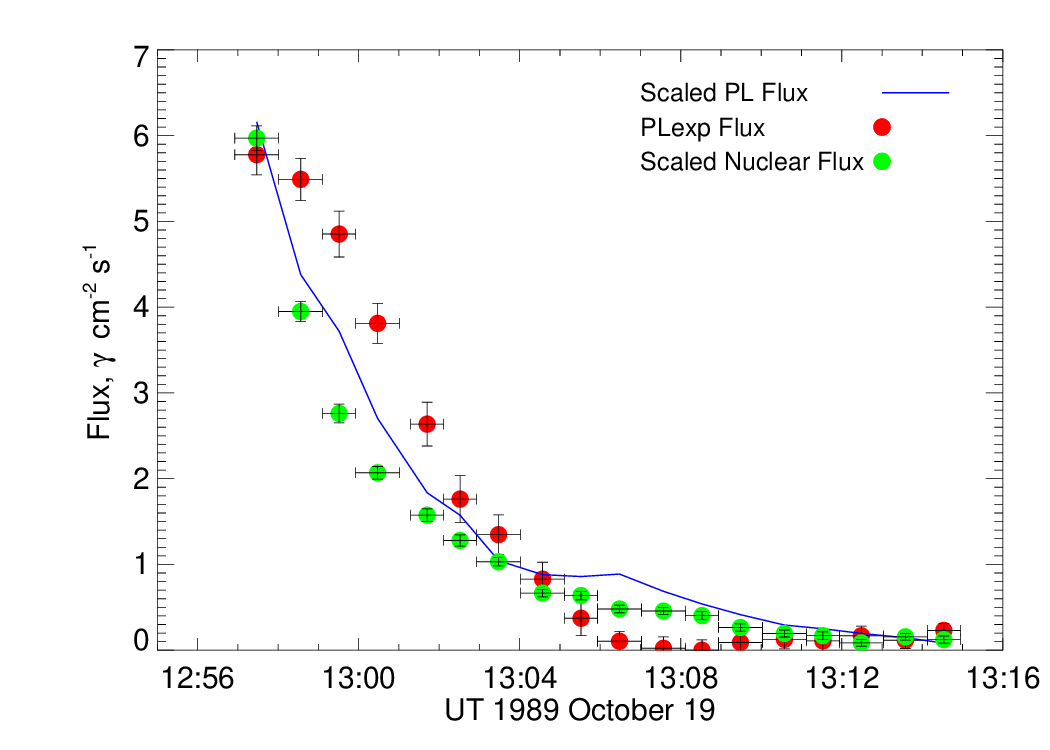}{0.5\textwidth}{(c)}
        \fig{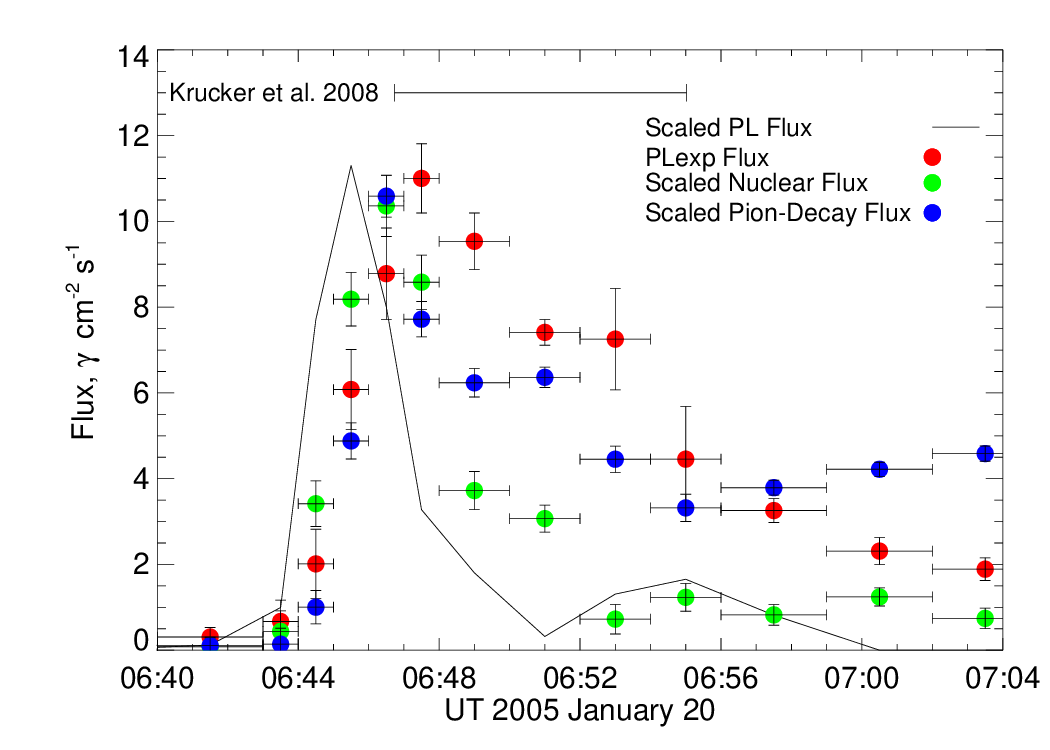}{0.5\textwidth}{(d)}}        
\caption{Temporal variations of the fitted power-law (PL), the hard Power-Law $\times$ exponential (PLexp), and the total nuclear de-excitation line fluxes $>$300 keV in three {\it SMM} and one {\it RHESSI} flares suggest a distinct origin of the PLexp component.  The $>$ 300 keV PLexp flux is shown on the Y-axis.  The  PL and nuclear-line fluxes have been scaled for comparison. The $\pi$-decay flux time history observed in the 2005 January 20 flare is different from all of the other histories. 
\label{timehist}}
\end{figure}

The statistical significance of the fluxes in most of the 27 flares are sufficient to compare the temporal variability of the PL, PLexp, and nuclear components.  We find that the PLexp component has a different time structure at some point in 18 of the flares.  Below, we discuss emissions in seven of the flares where the PLexp and PL time profiles are distinctly different.  In Figure \ref{timehist} we plot $>$300 keV fluxes for three GRS flares and one {\it RHESSI} flare.  Although the PLexp (red-filled circles) fluxes followed those of the PL (blue curves) and nuclear components (green-filled circles) early in the 1981 April 27 (panel (a)) and 1989 March 6 (panel (b)) flares, they were significantly weaker at later times. While the PL and nuclear fluxes were falling as observations began after the impulsive phase of the 1989 October 19 flare (panel (c)), the PLexp flux appeared to be rolling over from a peak and the fell rapidly relative to them after 13:04 UT.  Following the peak of 2005 January 20 flare (panel (d)), both the PL (black curve) and nuclear fluxes fell rapidly while the PLexp flux remained at a high-level for several minutes.  We also note the distinctly different $\pi$-decay (blue-filled circles) time history in that flare.  The significant increase in the PLexp/PL flux ratio following the impulsive phase of the January 20 flare is also clearly seen in three other {\it RHESSI} flares plotted in Figure \ref{212_th} of $\S$\ref{sec:disc}.  These significant differences in time histories suggest that the PLexp component has a distinct origin from the PL and nuclear components.

%\footnote{We note that the dominance of the PLexp component between 13:57 and 14:05 UT of the March 6 flare was during the same time interval when \citet{rieg98} reported three `electron-dominated' episodes.  During the first episode, beginning at 13:57 UT, we find that the exponential rollover energy, E$_R$, of the emission exceeded 10 MeV.  The emission was also observed in the 10--50 MeV band of the GRS high-energy matrix until 14:00 UT when E$_R$ decreased to $\sim$3 MeV.  Even though the early emission was dominated by the hard MeV continuum, the flare was a prolific accelerator of protons that produced nuclear $\gamma$ rays throughout.} 

\subsection{Location}\label{subsec:corona}

Our discussion above provides evidence that the PLexp emission is spectrally and temporally distinct from both the power-law hard X-ray emission and nuclear radiation in flares.  This suggests that it has a different origin than the emission from the electron and proton interactions occurring in the chromosphere.  \citet{kruc08} discovered a hard coronal source in their {\it RHESSI} imaging-spectroscopic measurements of the 2005 January 20 flare from 16:46:44 to 16:55 UT (its duration is denoted in Figure \ref{timehist}(d) by the solid black line).  It occurred after the PL and nuclear components had both decreased significantly and when the PLexp component dominated the MeV emission in our spatially-integrated observation. 

\begin{figure}[h!]
\centering
\includegraphics[width=130mm]{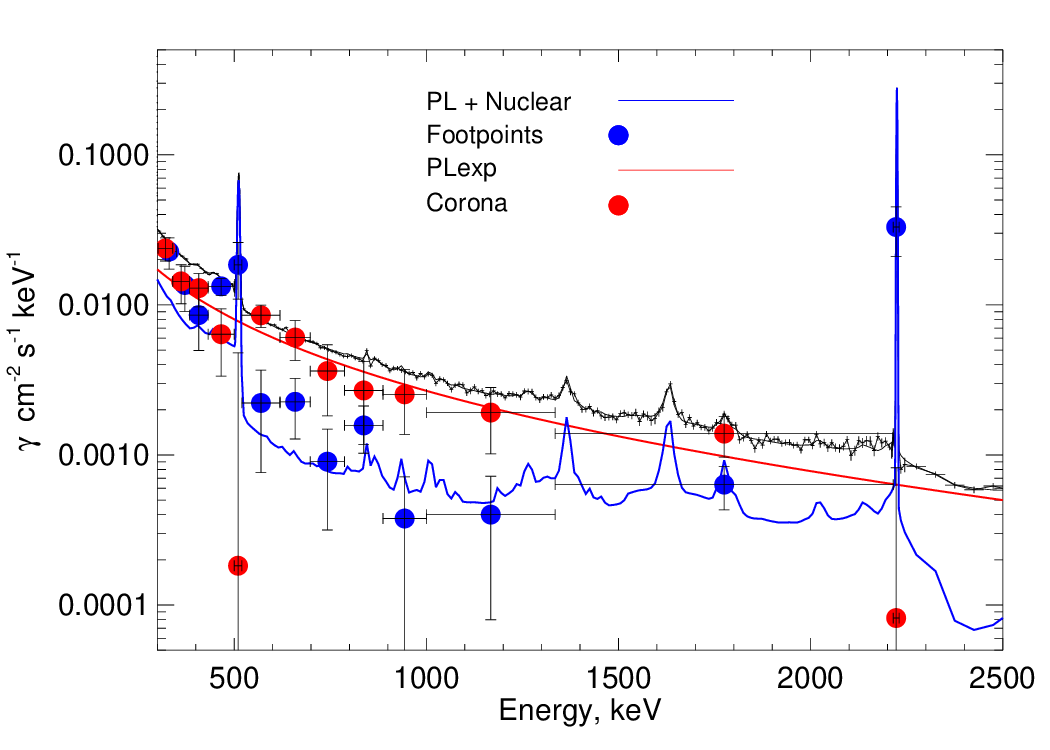} 
\caption{Spatially integrated spectrum (black curve) of the 2005 January 20 flare observed by {\it RHESSI} rear-detector segments with the fitted PL + nuclear (blue curve) and PLexp (red curve) components.    Footpoint (blue-filled circles; points at 320 and 361 keV have been shifted in energy for visibility) and coronal (red-filled circles) spectra were derived from imaging spectroscopy using {\it RHESSI's} rear detector segments.  The PLexp spectral component is consistent with the emission observed from the corona, while the PL and nuclear components are consistent with the emission from the footpoints.  
\label{05jan20_spec}}
\end{figure}

We have improved the {\it RHESSI} imaging spectroscopy for that time interval and extended its range to energies above 1 MeV.  In Figure \ref{05jan20_spec} we plot the derived $>$ 300 keV photon spectra of the radiation from the footpoint (blue-filled circles) and coronal (red-filled circles) sources from 16:46:44 to 16:55 UT.   Above 500 keV the coronal emission dominates with the exception of narrow energy bands containing footpoint 511 keV positron annihilation and 2.223 MeV neutron-capture lines.  The spatially-integrated photon spectrum (black curve) that we obtained during the same time interval is in good agreement with the sum of the spatially resolved footpoint and coronal spectra.   The spectrum of the fitted PLexp component (red curve) that was dominant during this time interval (Figure \ref{timehist}(d)) is in remarkably good agreement with the spatially-resolved coronal spectrum, indicating that this high-energy component originated in the corona.   The sum of the spatially integrated PL and line fluxes (blue curve) agrees well with emission from the footpoints, consistent with their origins deeper in the solar atmosphere.   Our studies using {\it RHESSI} front detector segments support these conclusions.  Although this is the only flare for which there are data to make this spatial and spectral comparison, it provides evidence that the PLexp emission is spatially distinct from the PL and nuclear radiations emanating from the footpoints, and that its location is consistent with being in the corona.  

\subsection{Directionality}\label{subsec:iso}

\begin{figure}[h!]
\centering
\includegraphics[width=80mm]{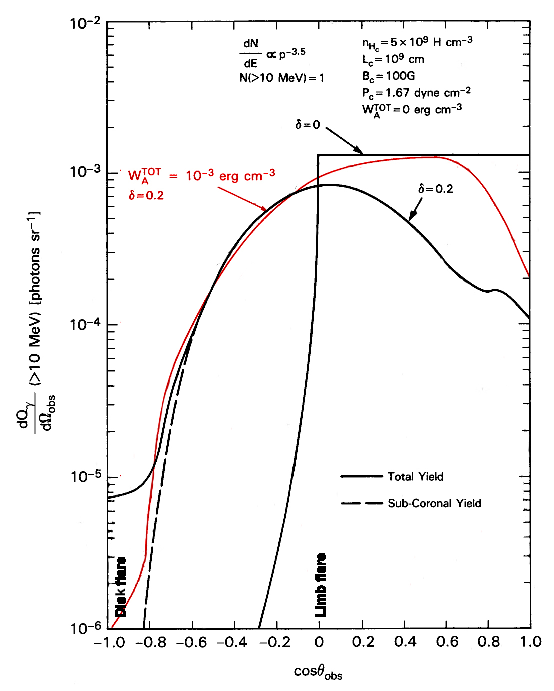} 
\caption{The angular distributions of the bremsstrahlung emission calculated by \citet{mill89} from a power-law momentum distribution of electrons injected into a magnetic loop of half length L$_c$ from the corona (density n$_h$, pressure P$_c$, magnetic field B$_c$) for magnetic convergence, $\delta = 0$,  and $\delta = 0.2$, with and without MHD turbulence W$_k$.  An observer of a flare at disk center would observe the flux at cos$\theta_{obs} = -1.0$ and an observer of a flare at the limb would observe the flux at cos$\theta_{obs} = 0$.  
\label{miller_angdist}}
\end{figure}

Bremsstrahlung and inverse Compton scattering by relativistic electrons can produce MeV $\gamma$ radiation.  These radiations are directed along the electron velocity vector and become more tightly beamed as the electron's energy increases (e.g. \citet{petr85,mack10}).  \citet{mill89} studied the bremsstrahlung produced by an isotropic distribution of $>$10 MeV coronal electrons injected into a magnetic loop.  The radiation pattern from inverse Compton scattering at these energies should be similar.   In Figure \ref{miller_angdist} we plot the calculated angular distributions of $\gamma$ rays for three different cases of magnetic convergence and turbulence.  For a disk flare, $\theta_{obs}$ = 180$^{\circ}$ and cos($\theta_{obs}$) = -1;  for a limb flare, ${\theta}_{obs}$ = 90$^{\circ}$ and cos($\theta$$_{obs}$) = 0.  The amount of magnetic convergence is related to the parameter $\delta$.  Below the transition region, the magnetic field strength is assumed proportional to P$^{\delta}$, where P is the pressure.  For coronal and photospheric pressures of 0.2 and 10$^5$ dyne cm$^{-2}$ and relevant magnetic field strengths of 100 and 1600 G, respectively, $\delta$ = 0.2.

For $\delta$ = 0, there is no convergence, and the interacting particle distribution and its high-energy radiation are downward isotropic with almost no radiation escaping from the Sun.  For $\delta$ = 0.2 and no turbulence, the angular distributions become fan beams peaking at 90$^\circ$ from the downward direction. The strongest escaping radiation is from limb flares (cos($\theta_{obs}$) = 0), as seen in Figure \ref{miller_angdist}, where the flux is two orders of magnitude more than that from a disk flare (cos($\theta_{obs}$) = -1. When significant turbulence is present in the loops, pitch-angle scattering precipitates electrons into the loss cone, increasing the number of downward-directed electrons. There is little change in the escaping radiation with flare location until the location nears disc center, where there is almost an additional order of magnitude decrease in the escaping flux.  This is consistent with the \citet{vest91} finding that the flares observed at energies above 300 keV showed a limb brightening.   In a study of the redshifts of nuclear lines as a function flare heliocentric angle \citet{shar02} found that the angular distribution of the interacting ions is consistent with that from pitch angle scattering in a converging magnetic loop.  In the presence of such scattering \citet{mill89} show in their Figure 13 that the $>$ 10 MeV energy spectrum does not change significantly for angles $>$ 90$^{\circ}$.  This means that one would not expect to observe any changes in the hardness of the spectrum of this high-energy radiation when viewing flares from disk center to the limb, but there would be an orders of magnitude change in intensity.

In contrast to these high-energy calculations, \citet{vest87} found that the $>$300 keV bremsstrahlung from flares hardened with increasing heliocentric angle.  This hardening for flares observed from disk center to the limb is due to the much broader angular distribution of bremsstrahlung from these lower-energy electrons and the fact that the photon energy decreases with increasing angle from the electron direction.   We see a clear example of this hardening with angle when we compare the flat PL spectrum (blue line) shown in Figure \ref{phspectra} (a) for a flare near the solar limb with the steep PL spectra in Figures \ref{phspectra}(b) \& (c) for flares near disk center.\footnote{It is surprising that the PL component of the 2002 July 23 flare plotted in Figure \ref{phspectra}(d) looks similarly steep even though the flare was at 73$^{\circ}$.  However,  \citet{smit03} found that the Doppler shifts of nuclear lines in that flare were in better agreement with a flare at a heliocentric angle of 30--40$^{\circ}$, suggesting that the flare loops were significantly tilted toward Earth.} 

\begin{figure}[h!]
\gridline{\fig{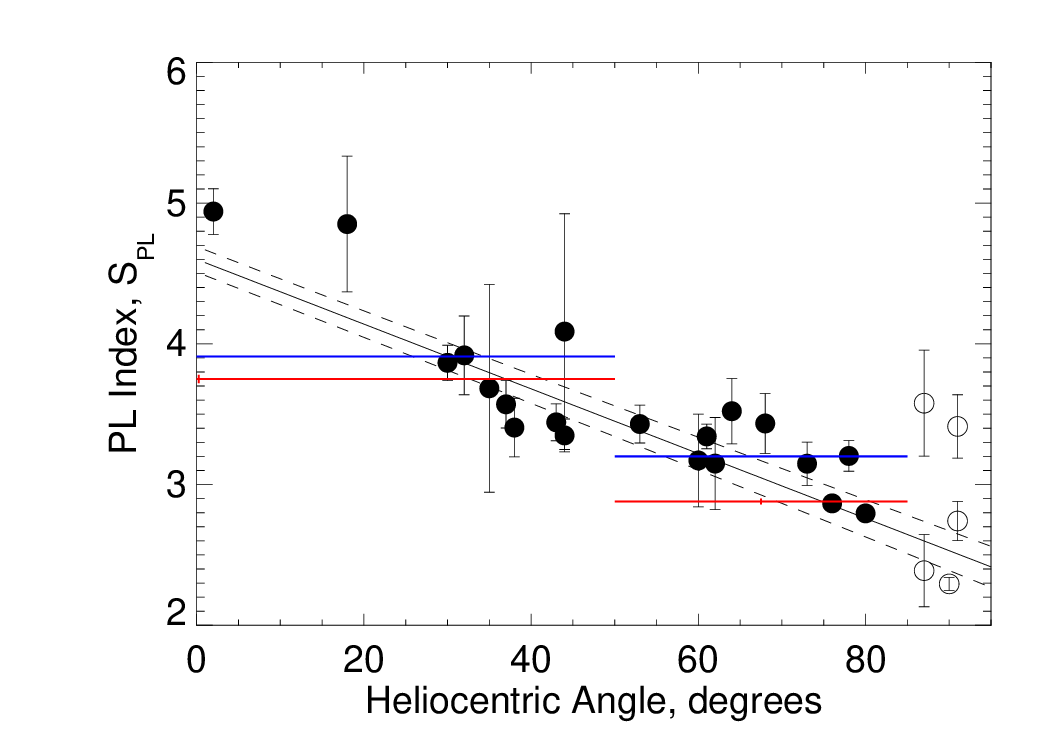}{0.5\textwidth}{(a)}
         \fig{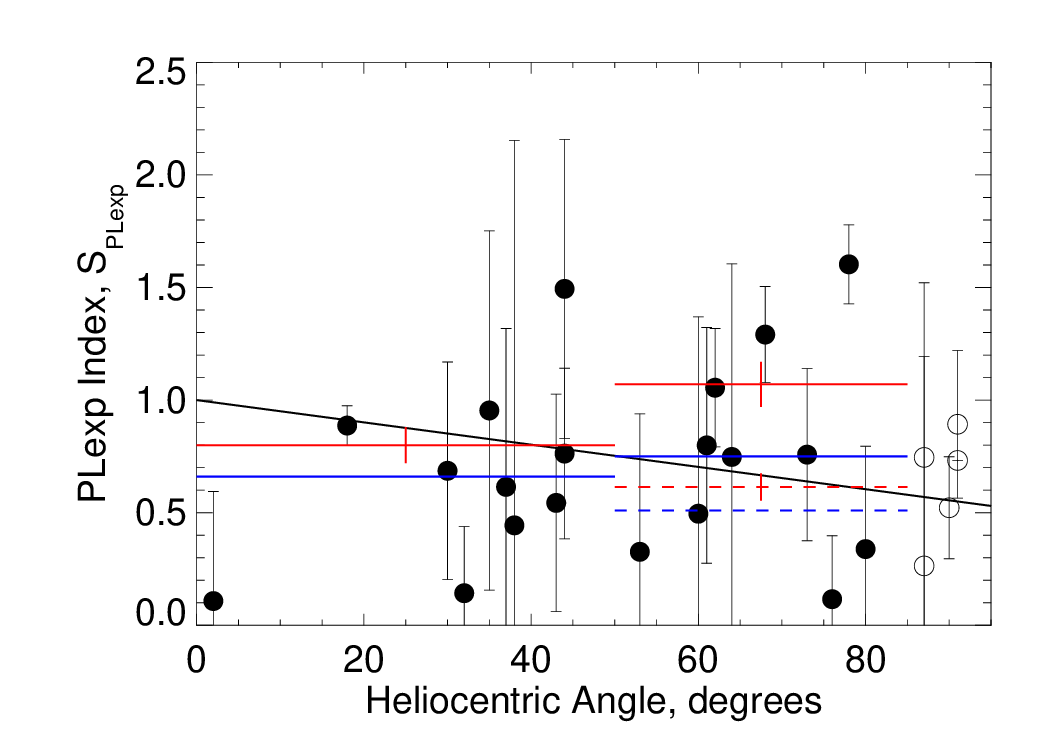}{0.5\textwidth}{(b)}} 
\gridline{\fig{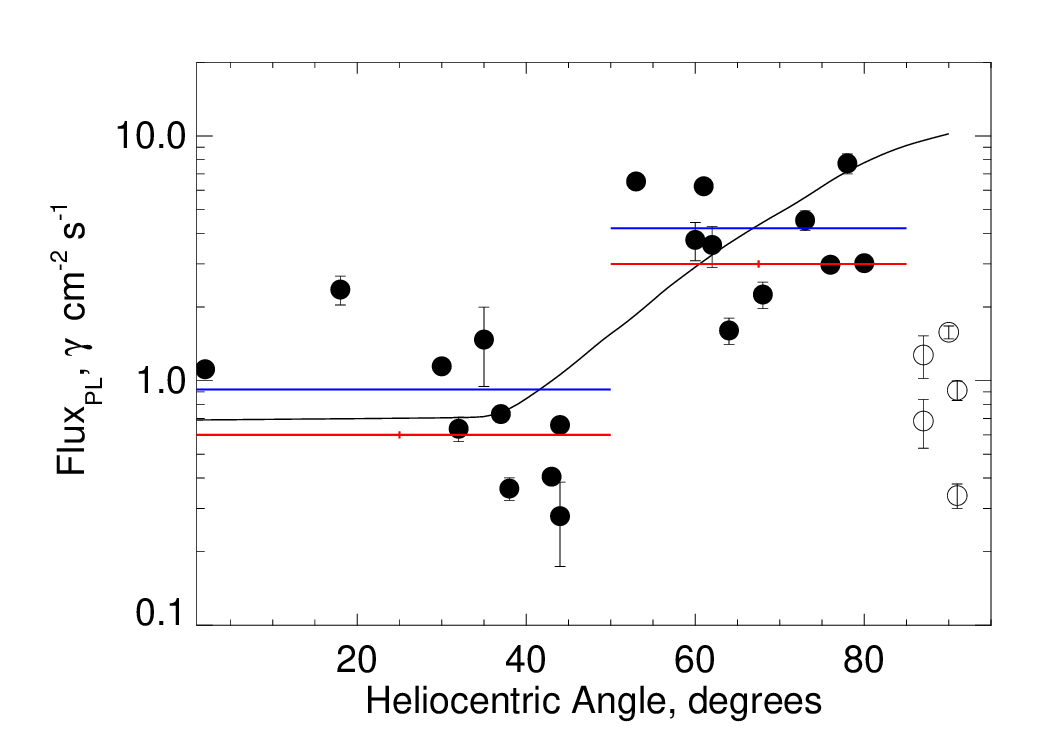}{0.5\textwidth}{(c)}
        \fig{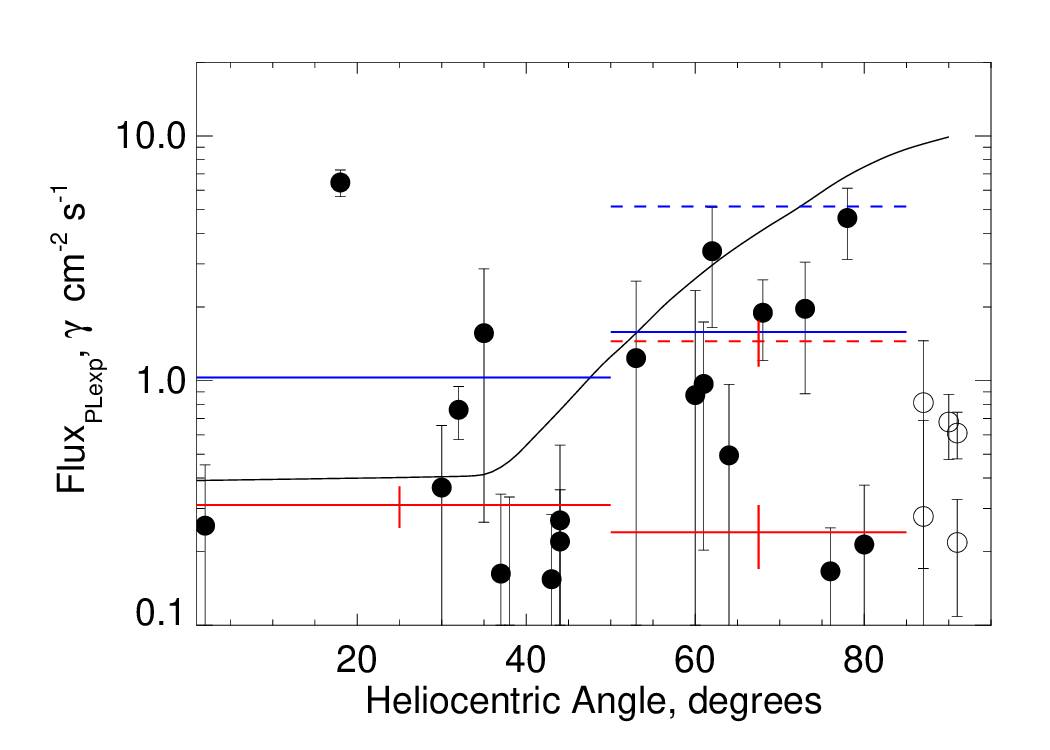}{0.5\textwidth}{(d)} }        
\caption{Panels (a) and (b) show the heliocentric angle dependence of the PL index, S$_{PL}$, and PLexp index, S$_{PLexp}$. The best fit S$_{PL}$ linear variation and $\pm 1 \sigma$ uncertainties is shown by the solid and dashed lines, respectively in panel (a).  This variation is shown by the solid line in panel (b).  Panels (c) and (d) show the heliocentric angle dependences of Flux$_{PL}$ and Flux$_{PLexp}$.  The heliocentric variation for bremsstrahlung from $>$ 10 MeV electrons calculated by \citet{mill89} is plotted as solid curves in both panel.   Means (solid blue lines) and weighted means (solid red lines with uncertainties) of the measured values from 0$^{\circ}$ to 50$^{\circ}$ and 50$^{\circ}$ to 85$^{\circ}$ are plotted in all four panels. The dashed lines from 50$^{\circ}$ to 85$^{\circ}$ in panels (b) and (d) are the estimated means (blue) and weighted means (red) based on the variation observed in panels (a) and (c), respectively. We restricted our fits to angles $<$85$^{\circ}$ (filled circles) due to attenuation effects in flares near the limb (open circles).  The high fluxes near 18$^{\circ}$ came from the intense 2003 October 28 flare that was observed by RHESSI after the impulsive phase. 
\label{helang}}
\end{figure}

Details of the angle-dependent characteristics of the PL and PLexp components are revealed using data from Table \ref{tab:sum}.  The significant decrease in the PL index, S$_{PL}$, $>$ 300 keV plotted in Figure \ref{helang}(a) reveals the strong anisotropy of the electrons in the turbulent magnetic loops.  Our linear fit to the index \textit{vs} heliocentric angle for flares at angles $<$85$^{\circ}$ is shown by the solid line with the dashed lines showing the uncertainties. S$_{PL}$ decreases from 4.6 $\pm$ 0.1 at disk center to 2.4 $\pm$ 0.1 at the limb. This decrease is more significant than the 3.4 to 2.7 index change measured by \citet{vest87}.  The lower index found by \citet{vest87} near disk center is due to the hard nuclear-line and PLexp continua that were not accurately subtracted from the spectra.  The large uncertainties in S$_{PLexp}$ plotted in Figure \ref{helang}(b) are due to the strong PL continuum which masks the PLexp component at low energies.  A reduction by about a factor of two in S$_{PLexp}$ from disk center to the limb, as found for the PL (solid curve), appears unlikely but is masked by the uncertainties.  

In order to determine if S$_{PLexp}$ shows the same decrease with heliocentric angle as S$_{PL}$, we plot mean (solid blue) and weighted mean (solid red) indices for heliocentric angles from 0$^{\circ}$ to 50$^{\circ}$ and from 50$^{\circ}$ to 85$^{\circ}$ in panels (a) and (b).  For comparison in panel (b) we also plot the expected mean and weighted mean values of S$_{PLexp}$ for heliocentric angles from 50$^{\circ}$ to 85$^{\circ}$ (dashed lines) assuming they show the same decrease as S$_{PL}$.  The measured weighted mean value of 1.07 $\pm$ 0.10 for S$_{PLexp}$ is inconsistent with the expected value of 0.61 $\pm$ 0.06 with $>$ 99.7\%  confidence.  However, because \citet{mill89}'s calculations for $>$ 10 MeV electrons did not show spectral variation with heliocentric angle, it is likely that the expected change in PLexp index would not be as large.  The variation of S$_{PLexp}$ appears to be consistent with an isotropic distribution. We note that the weighted mean of S$_{PLexp}$ is $<$ 1 and some values are significantly below 1, which is inconsistent with a bremsstrahlung origin.  We discuss this issue in $\S$\ref{sec:plexp_par} and in $\S$\ref{sec:origin}, but note that OSPEX may not calculate the uncertainties accurately when the parameters are strongly correlated (see $\S$\ref{sec:fits} and \citet{irel13}).

From the work of \citet{mill89} and \citet{vest87a}, a better test for anisotropy of the PLexp component is how its flux varies with heliocentric angle.  For an anisotropic distribution of 300 -- 350 keV bremsstrahlung photons, \citet{vest87a} estimated that fluxes of flares observed near the solar limb would be about a factor seven higher than if they were located near disk center.  We plot the measured values of Flux$_{PL}$ {\it vs} heliocentric angle in Figure \ref{helang}(c).  The large spread in fluxes reflects the flare-to-flare variations such as seen in the intense 2003 October 28 flare observed near a heliocentric angle of 18${^\circ}$.  Even with this spread in flux values, there appears to be a significant increase in Flux$_{PL}$ for flares at large heliocentric angles that is comparable to the estimated variation from the work of \citet{mill89} that is plotted as a solid curve. Both the mean (solid blue line) and weighted mean (solid red line) fluxes for heliocentric angles from 50$^{\circ}$ to 85$^{\circ}$ are about 5 times higher than the means for angles from 0$^{\circ}$ to 50$^{\circ}$. In contrast, there is no significant change in the mean and weighted mean PLexp fluxes for the two ranges of heliocentric angles plotted in Figure \ref{helang}(d).  The weighted mean Flux$_{PLexp}$ is 0.31 $\pm$ 0.06 for 0$^{\circ}$ to 50$^{\circ}$ and 0.24 $\pm$ 0.07 for 50$^{\circ}$ to 85$^{\circ}$.  The dashed blue and red lines show the expected mean and weighted mean (1.45 $\pm$ 0.31) fluxes from 50$^{\circ}$ to 85$^{\circ}$ had they increased by the same factors as found for Flux$_{PL}$ plotted in panel (c).   The heliocentric variation in Flux$_{PLexp}$ is consistent with an isotropic distribution and is inconsistent with the large change in Flux$_{PL}$ \textit{vs} heliocentric angle with $>$ 99.7\% confidence.

From the work of \citet{mill89}, electrons producing such an isotropic distribution would likely radiate from the corona and not from the footpoints.  This is consistent with the coronal location of the PLexp component in the 2005 January 20 flare. 

\section{Spectral Characteristics of the PLexp Component}\label{sec:plexp_par}

Three parameters define the PLexp component: its power-law index, S$_{PLexp}$, its rollover energy, E$_R$, and it flux relative to the PL component, Flux$_{PLexp}$/Flux$_{PL}$.  Table \ref{tab:sum} lists the parameters and their uncertainties that provide these values in the 25 strong nuclear line flares that we studied.  In order to expand the dynamic range of our study we included `electron-dominated' episodes \citep{rieg98}, discussed in Appendix \ref{sec:rieger}, and weak solar flares, discussed in Appendix \ref{sec:weak}.  These events extended the dynamic range of the flare fluxes that we studied to almost three orders of magnitude (Appendix \ref{sec:eicorr}). 

%In order to be more comprehensive, we expanded our study to include `electron-dominated' episodes \citep{rieg98} and weak flares. In Appendix \ref{sec:rieger} we discuss our study of the episodes

The measured indices of the PLexp component, S$_{PLexp}$, in the 25 large flares are displayed in Figure \ref{helang}(b) and their weighted mean is 0.80 $\pm$ 0.08.  Thus the mean index is consistent with a value $<$ 1 with 98\% confidence.\footnote{We find that the same is true for `electron-dominated' episodes.  The fitted index for weak flares is 0.5 $\pm$ 0.4.  Thus the PLexp index is consistent with values of $\lesssim$ 1 for flares over three orders of magnitude in intensity.}  If the PLexp emission is due to bremsstrahlung, S$_{PLexp}$ cannot be $<$ 1, but its measurement and uncertainties are compromised by the presence of the strong PL component and the strong correlation of the parameters discussed at the end of $\S$\ref{sec:fits}.   Early in the 2005 January 20 flare, between 06:44 and 06:46:44 UT, when the PL component was strong, we measured an S$_{PLexp}$ value of 1.0 $\pm$ 0.3.  However, later in the flare from 06:46:44 to 06:55 UT, during the time when we imaged the source and when the PL component was weak, we measured a value of 1.39 $\pm$ 0.06, consistent with a bremsstrahlung origin.  In $\S$\ref{subsec:thin2} we find that the spectra in the 17 flares studied can be fit with thin-target bremsstrahlung instead of the PLexp function with a $>$ 85\% probability.  Thus, a bremsstrahlung origin for the PLexp component cannot be ruled out.

\begin{figure}[h!]
\gridline{\fig{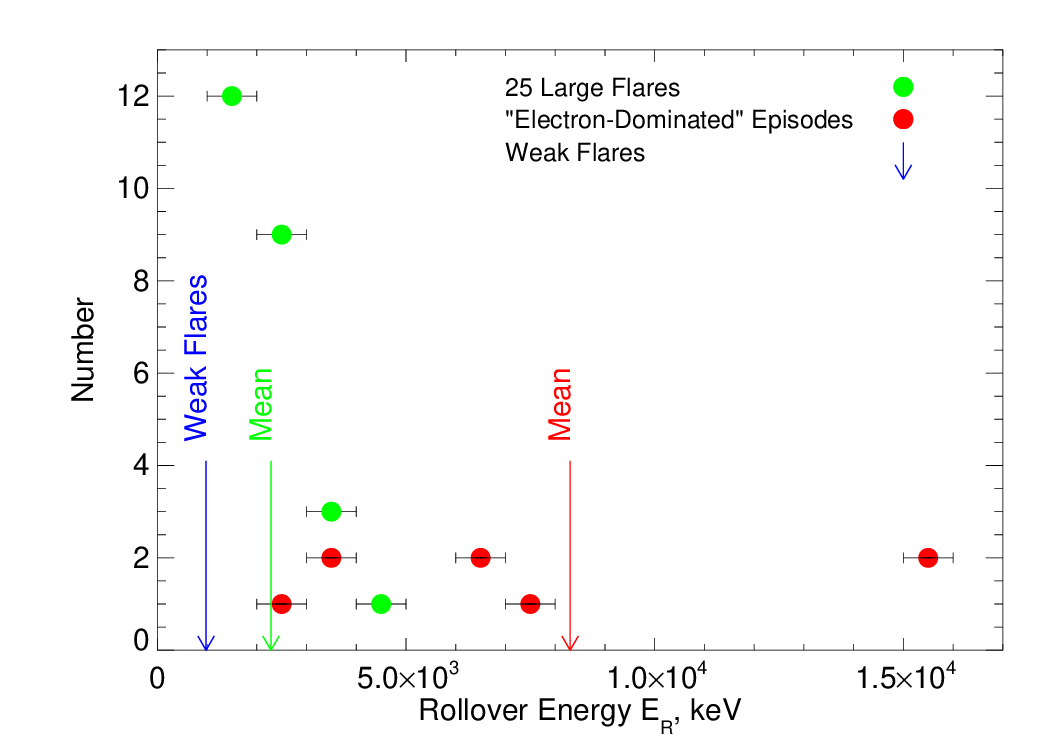}{0.5\textwidth}{(a)}
         \fig{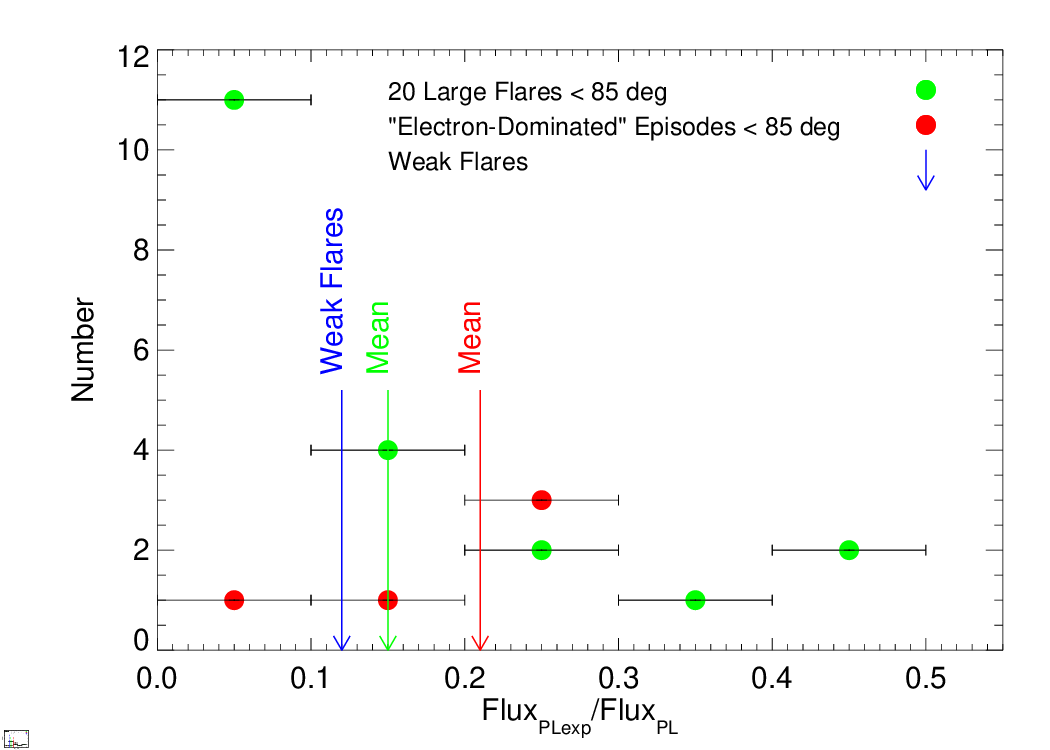}{0.5\textwidth}{(b)}} 
    \caption{Panel (a): Number distributions of the exponential rollover energy (E$_R$) for 25 nuclear-line flares (green filled circles) and 8 `electron dominated' episodes \citet{rieg98} (red filled circles). Panel (b): Number distribution of the Flux$_{PLexp}$/Flux$_{PL}$ ratio $>$ 0.3 MeV, after correcting the anisotropic Flux$_{PL}$ for heliocentric angle, for 20 flares and 5 `electron dominated' episodes occurring at heliocentric angles $<$ 85$^{\circ}$.   The green and red vertical arrows mark the mean values for the large flares and `electron dominated' episodes, respectively.  The blue vertical arrows denote the fitted values for the sum of 48 weak flares.
\label{distributions}}
\end{figure}

The number distribution of rollover energies, E$_R$, for large nuclear-line flares, plotted as filled green circles in Figure \ref{distributions}(a), peaks at $\sim$ 1.5 MeV, extends up to $\sim$ 5 MeV, and has a mean of 2.2 MeV.  In contrast, the E$_R$ distribution of the `electron-rich' episodes is much flatter with a mean value four times higher.  For weak flares, the fitted rollover energy of 890 $\pm$ 40 keV (blue arrow) is less than any energy found in large flares.  

In Figure \ref{distributions}(b) we plot the number distribution of PLexp to PL flux ratios $>$ 300 keV, Flux$_{PLexp}$/Flux$_{PL}$.  We corrected the anisotropic Flux$_{PL}$ to a heliocentric angle of 80$^{\circ}$ using a fit to the data plotted in Figure \ref{helang}(c).  Corrected to this heliocentric angle, the radiation from a broad downward or fan beam angular distribution of electrons would be effectively isotropic.  We restricted our study to flares $<$85$^\circ$ to avoid effects due to attenuation of the PL footpoint emission at the limb.  For large flares, the Flux$_{PLexp}$/Flux$_{PL}$ distribution peaks near 0.05, extends up to $\sim$ 0.5, and has a mean of 0.15.  Thus the $>$ 300 keV flux in the PLexp component is typically less than $\sim$ 20\% of the $>$ 300 keV flux in the PL component for large flares. The Flux$_{PLexp}$/Flux$_{PL}$ distribution for the five `electron-rich' episodes at heliocentric angles $<$ 85$^{\circ}$, red-filled circles, is flat and has a mean value $\sim$ 40\% higher than that in the large flares.  The Flux$_{PLexp}$/Flux$_{PL}$ ratio in weak flares (blue arrow) is only $\sim$ 20\% lower than the mean value found for large flares.

Thus, the rollover energy, E$_R$, exhibits the most significant difference, varying from a low of $\sim$ 0.9 MeV in weak flares to a mean value of 2.2 MeV in nuclear-line flares and to significantly higher values in `electron-dominated' flares. The Flux$_{PLexp}$/Flux$_{PL}$ ratio follows the same increasing trend but not as strikingly. 

%In Figure \ref{distributions}(b) we plot the number distribution of PLexp to PL flux ratios $>$ 300 keV, Flux$_{PLexp}$/Flux$_{PL}$, for flares at heliocentric angles $<$85$^\circ$.  To compare with the isotropic PLexp component we corrected the anisotropic Flux$_{PL}$ to a heliocentric angle of 80$^{\circ}$ using the empirical function plotted in Figure \ref{helang}(c).  Corrected to this heliocentric angle, radiation from a broad downward angular distribution of electrons would be isotropic.  We restricted our study to flares $<$85$^\circ$ to avoid possible strong attenuation of the PL footpoint emission at the limb.
 
%Thus, the most significant difference in the PLexp component for weak flares relative to strong flares is its low $\sim$ 900 keV rollover energy.  In contrast, the typical rollover energy for `electron-rich' episodes is 2-4 times higher than in strong flares.}

\section{Electron Spectrum Producing the MeV Gamma-Ray Continuum}\label{sec:origin}

We have identified a solar-flare MeV $\gamma$-ray continuum that is temporally distinct from both the power-law extension of the hard X-ray emission and nuclear emission.   We have provided evidence that the continuum is consistent with being isotropic and thus not likely to originate in the footpoints. In fact, imaging spectroscopic measurements during the 2005 January 20 flare indicates that it originates in the corona.  We have represented its spectrum by a power law times an exponential, PLexp (see Equation 1), where the index of the power-law is typically $\lesssim$ 1 and the exponential rollover energy, E$_R$, ranges from about 1 to 5 MeV in large nuclear line flares.   There are two primary mechanisms by which electrons can produce MeV $\gamma$ rays: inverse Compton scattering and bremsstrahlung.  Below we discuss these processes, how well they are able explain the features of the PLexp component, and the characteristics of the electron spectra that would explain the observations.

\subsection{Electron Compton Scattering of Flare X-rays}\label{subsec:compton}

\begin{figure}[h!]
\gridline{\fig{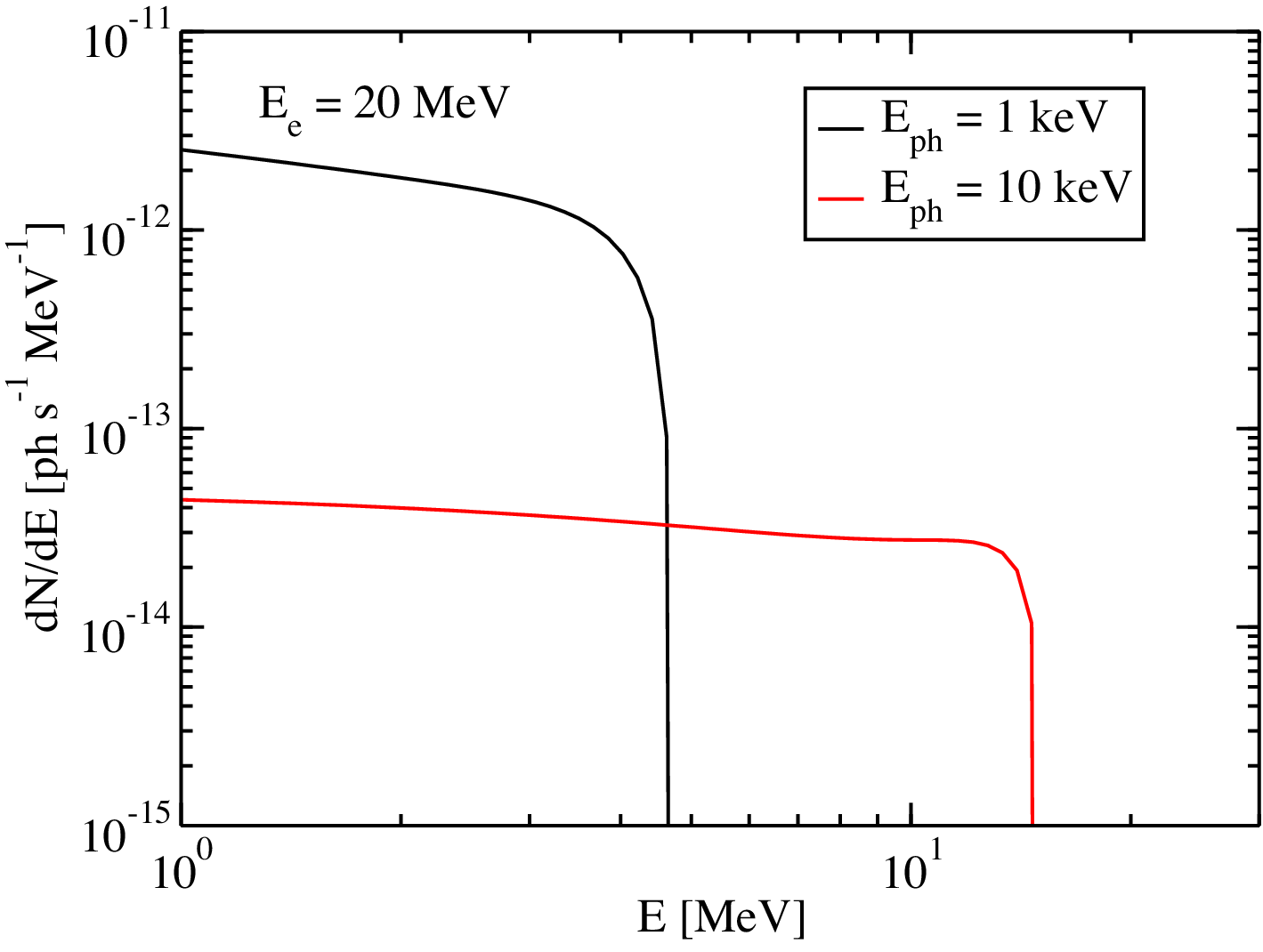}{0.45\textwidth}{(a)}
         \fig{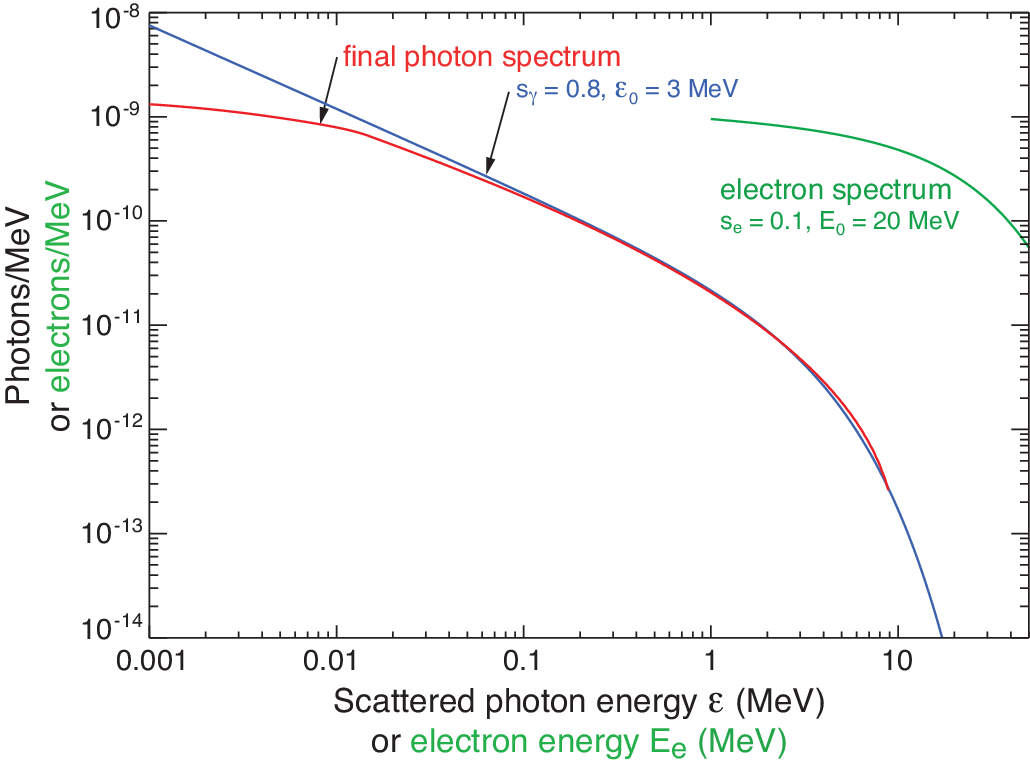}{0.45\textwidth}{(b)}}       
\caption{ (a) Spectra produced when isotropic 20 MeV electrons Compton scatter an isotropic population of 1 and 10 keV X-rays to MeV energies. (b) An electron spectrum in the shape of a power law with index 0.1 and exponential rollover energy of 20 MeV (green curve) will produce a $\gamma$-ray spectrum (red curve) consistent with a PLexp with index of 0.8 and 3 MeV rollover energy (blue curve). 
\label{comp}}
\end{figure}

Flat flare spectra at hundreds of keV energies, such as we observe in the PLexp component, prompted researchers to consider inverse Compton scattering as an alternative to bremsstrahlung \citep{kruc08b}.   Such emission is expected to dominate at low coronal mass densities and high-electron energies.  \citet{mack10} made detailed calculations of the emission produced when electrons and positrons with energies reaching 100 MeV up-scattered optical and EUV photons.  \citet{chen12} refined these calculations and extended them to include scattering of soft X-rays emitted during flares.  Using the work of \citet{jone68} and \citet{blum70}, we calculated $\gamma$-ray spectra produced when isotropic distributions of high-energy electrons Compton scatter off isotropic photon distributions.  In Figure \ref{comp}(a) we plot the spectrum produced when 20 MeV electrons scatter 1 and 10 keV X-rays.  We see that 20 MeV electrons Compton scatter 1 keV X-rays to produce a flat spectrum with a sharp rollover above 4 MeV.  A photon spectrum that better reflects the shape of the PLexp component that we observe in nuclear-line flares would be produced by an electron distribution of the form
\begin{flalign}
n(E_e) \propto E_e^{-s_e} \exp{-(E_e/E_0)}\ .
\end{flalign}
where n is the number flux of electrons with energy E$_e$, s$_e$ is the index of the power-law function and E$_0$ is the exponential rollover energy. An example of such an electron spectrum with s$_e = 0.1$ is plotted in Figure \ref{comp}(b) along with the resulting photon spectrum that would be produced by inverse Compton scattering of soft X-rays.  This photon spectrum is consistent with a PLexp spectrum with an index $<$ 1, which raised questions about a bremsstrahlung origin that we discussed in $\S$\ref{sec:plexp_par}.  It is beyond the scope of this paper for us to fit the data to obtain the best electron spectral distribution, but we expect that spectrum to rollover at energies below 20 MeV.

\subsection{Electron Thin-Target Bremsstrahlung}\label{subsec:thin2}
Here we determine whether thin-target bremsstrahlung can produce the flat spectrum represented by the PLexp component, and determine the characteristics of the electron spectrum.  We use the SSWIDL OSPEX `thin2' function to fit the MeV continuum.  Unfortunately, only electron-ion (e-i) bremsstrahlung is incorporated in `thin2'.  \citet{kont07} showed that an upward break (change of $\sim$0.4 in index) in the power-law spectrum in the 2005 January 17 flare can be explained by including electron-electron (e-e) bremsstrahlung.  However, this amount of hardening is too small to account for the PLexp component discussed in this paper.  We find that including e-e bremstrahlung in `thin2'  would decrease the estimated electron flux by about a factor of two but has a less significant impact on determination of the shape of the electron spectrum (\citet{haug98,opar20} and Ivan Oparin [2024, private communication]).  The `thin2'  function assumes that the electrons have a broken power-law energy spectrum with the normalization factor (electron flux $>$ 300 keV $\times$ source density $\times$ source volume), the break energy, and the indices below and above the break energy as free parameters.  

We were able to fit the spectra and obtain well-defined parameters in 17 nuclear-line flares using the `thin2' function and the same power-law (PL) and nuclear-line photon components used for the PLexp fits in $\S$\ref{sec:fits}.   Thus, bremsstrahlung is a viable origin for the hard PLexp component even though the spectral fits give a mean index $\lesssim$ 1 in our studies in $\S$\ref{subsec:iso}.  In our thin-target fits to the 2003 October 28 flare spectrum, the PL index has a 60\% correlation with the bremsstrahlung flux and a 40\% correlation with the index below the break energy.  There is no significant correlation between the PL amplitude and any of the thin-target parameters.   We have compared the quality of fits for the thin-target origin with those obtained in $\S$\ref{sec:fits} for the PLexp component.  Thin-target spectral fits have a mean probability that is 87\% of PLexp fits.  The probabilities range from 0.07 to 1.77.  It is surprising that the worst thin-target fit is for the 2014 February 25 flare that has the steepest PLexp index.  Our studies indicate that uncertainties in the PL component allow for thin-target emission in spite of the fitted hardness of the PLexp component (also see $\S$\ref{sec:plexp_par}).  When using thin-target bremsstrahlung in the fits instead of the PLexp the PL spectral index softened by 0.06 $\pm$ 0.02 and its flux decreased by 3.3 $\pm$ 1.0 \%. Thus the presence of the strong PL component affects the determination of the PLexp index as we found in $\S$\ref{sec:plexp_par} where the index had a value of 1.0 $\pm$ 0.3 when the PL intensity was strong and a value of 1.39 $\pm$ 0.06 when it was weaker.

\begin{figure}[h!]
\gridline{\fig{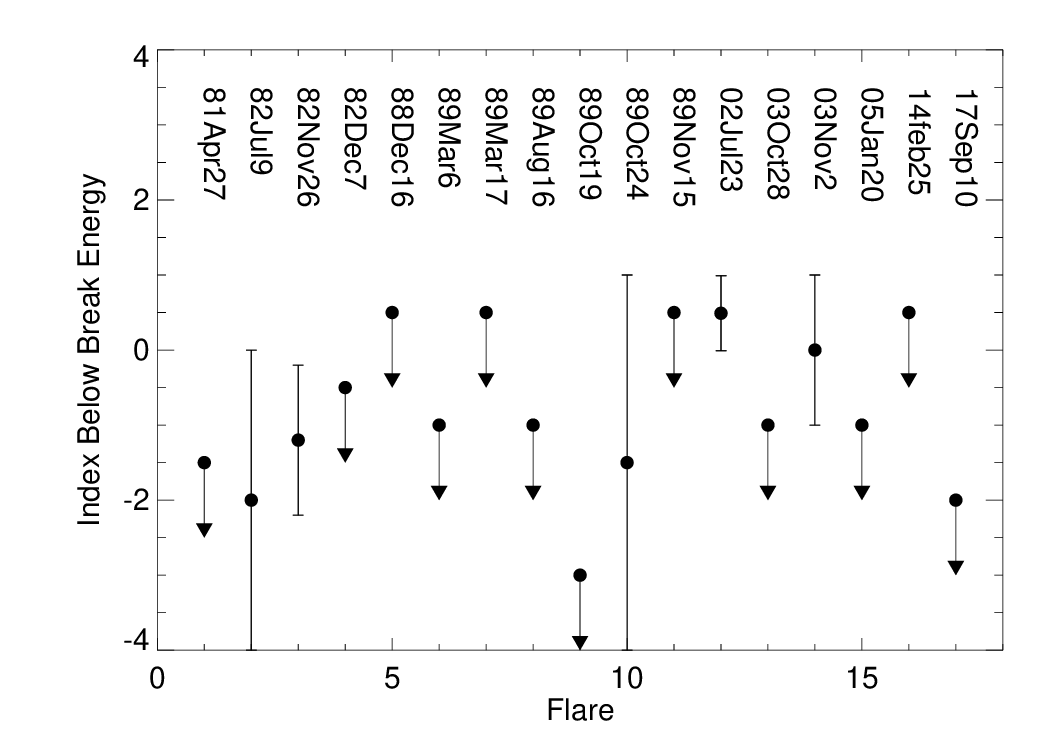}{0.5\textwidth}{(a)}
         \fig{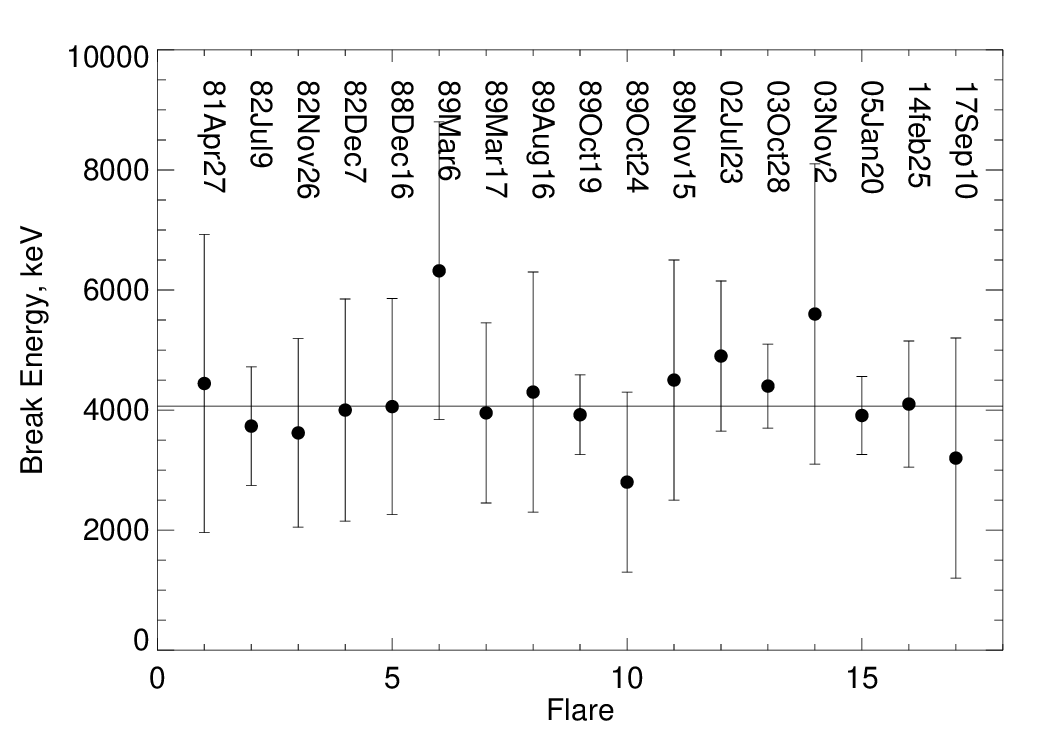}{0.5\textwidth}{(b)}}
\gridline{\fig{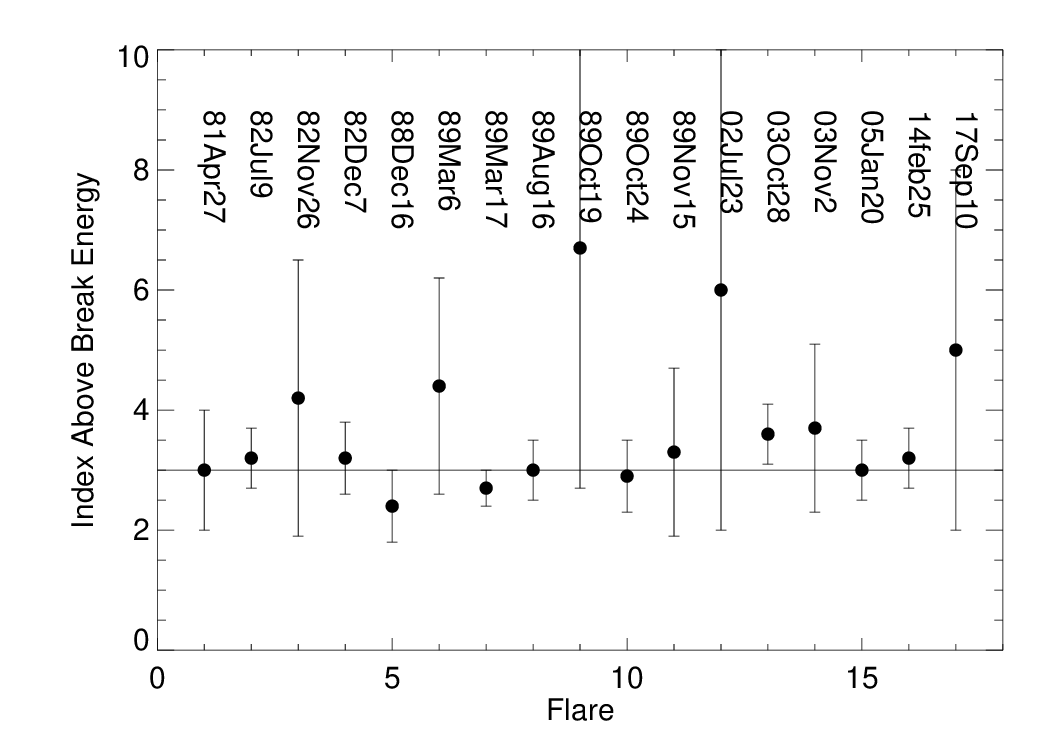}{0.5\textwidth}{(c)}
         \fig{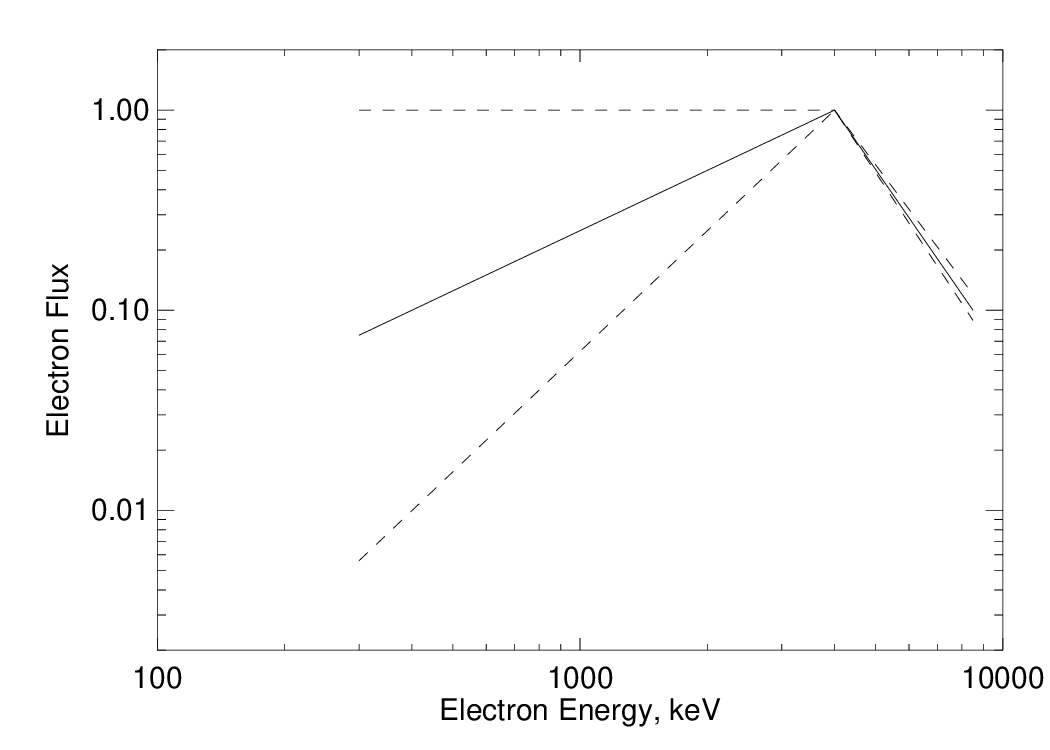}{0.5\textwidth}{(d)}}        
\caption{ Spectral characteristics of the new electron population in 17 nuclear-line flares assuming that the PLexp continuum is produced by thin-target bremsstrahlung.  The modeled electron spectrum has the shape of a broken power law.  Panel (a) shows the fitted indices below the break energy, panel (b) shows the fitted break energies, and panel (c) shows the fitted indices above the break energy. Panel (d) shows a representation of the broken power-law electron spectrum for this model with a mean 4 MeV break energy.  Break energies in the fitted flare spectra range between 3 -- 5 MeV and the dashed lines show the uncertainties in the fitted indices.   
\label{elspectra}}
\end{figure}

While the electron break energy and index above that energy were well determined in all of the thin-target fits, we needed to perform a $\chi^2$ minimization analysis to reliably constrain the index below the break energy.  We did this by manually changing its value while allowing all the other parameters to vary.  We plot the results of these fits in Figure \ref{elspectra}.   All of the power-law indices below the break energy in panel (a) are consistent with rising or flat electron spectra, while all of the power-law indices above the break energy are consistent a falling spectrum with a weighted mean index of 3.0 $\pm$ 0.2.  Given the large uncertainties, all of the break energies in panel (b) are consistent with their weighted mean of 4.1 $\pm$ 0.3 MeV.  Although a mono-energetic distribution of electrons can fit the data in a few flares, it generally produced significantly worse fits than a peaked distribution because the observed spectra require photons above the electron break energy.

In Figure \ref{elspectra}(d) we plot a representation of the MeV electron spectrum based on the fitted parameters plotted in the other three panels of the figure.  The spectrum is flat or rises to a peak energy near 4 MeV and then falls with a power-law index of about 3 at higher energies.  We also fit the summed spectrum of 48 weak flares with thin-target bremsstrahlung.  The average break energy for these flares is consistent with 4 MeV, the spectral index of 0.5 $\pm$ 0.5 below the break energy is comparable to the steepest observed in the large flares, and the power-law index of 6 $\pm$ 2 above the break energy is steeper than the mean in the large flares.

\section{Summary and Discussion}\label{sec:disc}

Although solar-flare hard X-ray emission has been well studied, understanding the electron-produced $\gamma$-ray emission at energies $>$ 300 keV has been more difficult.  One of the obstacles has been the contribution from nuclear $\gamma$ rays.  In a study of 25 nuclear-line flares, significant improvements in our understanding of both nuclear $\gamma$-ray line production and instrument performance have allowed us to reveal a distinct flat MeV $\gamma$-ray continuum  represented by a power law times an exponential (PLexp) function.  The spectral fits reliably separate this component from the power-law (PL) extension of hard X-ray emission that dominates solar spectra at lower energies and from the nuclear $\gamma$-ray emission at higher energies. 

%In this paper we use improved cross sections for producing these $\gamma$ rays and improved detector response matrices in a study of 25 nuclear-line flares to reveal a flat MeV $\gamma$-ray continuum well described by a power-law times an exponential (PLexp) function. 

\begin{figure}[h!]
\gridline{\fig{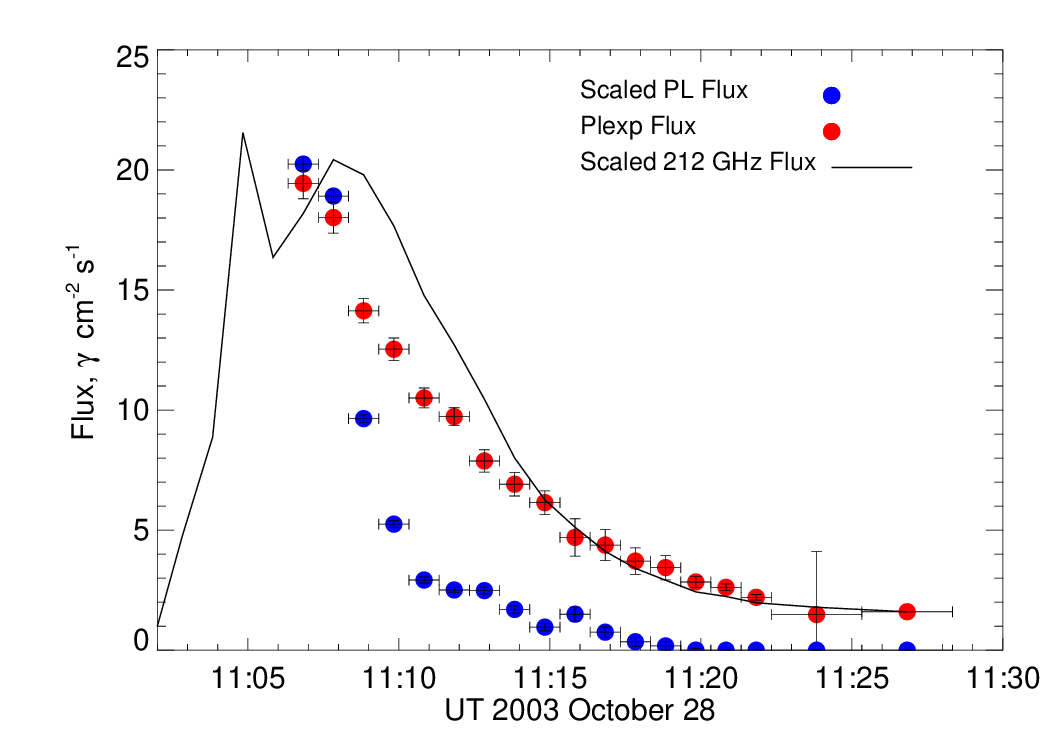}{0.5\textwidth}{(a)}
         \fig{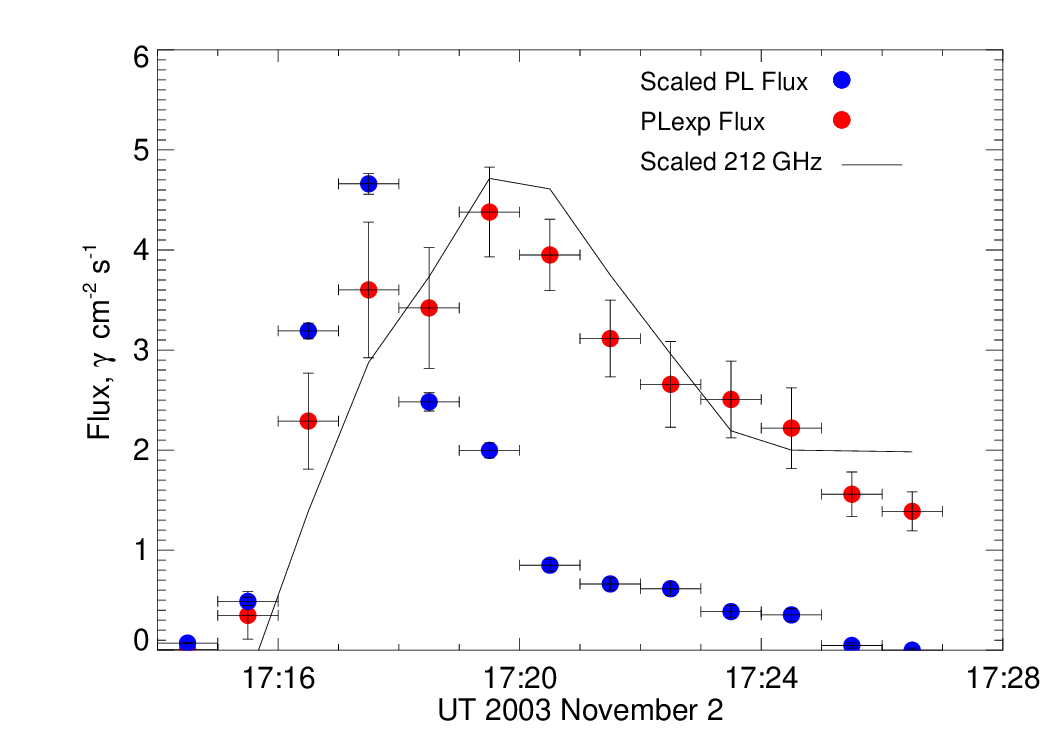}{0.5\textwidth}{(b)}}
\gridline{\fig{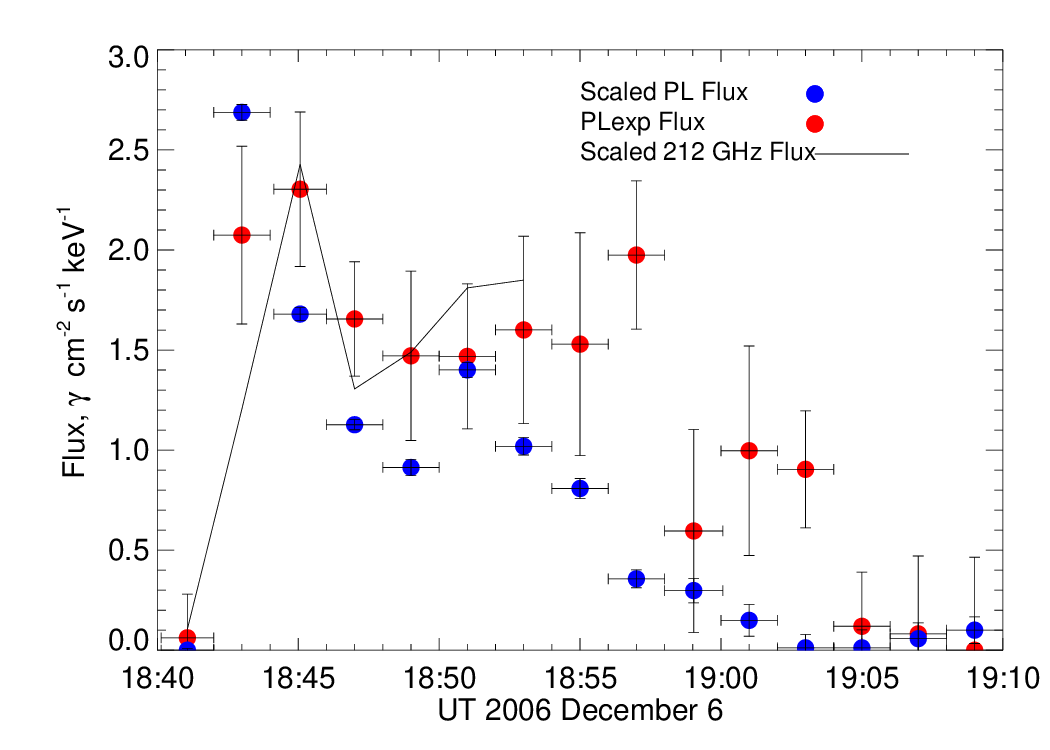}{0.5\textwidth}{(c)}}
\caption{Comparison of the time histories of the PL and PLexp fluxes in $\gamma$-ray spectra and the submillimeter fluxes in the 2003 October 28 \citep{luth04}
, 2003 November 2 \citep{silv07}, and 2006 December 6 \citep{kauf09} solar flares observed by {\it RHESSI}.
\label{212_th}}
\end{figure}

While the time history of the nuclear-line flux typically follows that of the $>$ 300 keV PL emission with a small delay, the time history of the PLexp component shows significant differences in 18 of the intense solar flares in our study.  We plot these differences for seven of the flares in Figures \ref{timehist} and \ref{212_th}.   This suggests that the PLexp emission may originate from a different source of accelerated electrons than the hard X-rays.  During the time interval in the 2005 January 20 flare when the PLexp emission was dominant,  \citet{kruc08} used {\it RHESSI} imaging spectroscopy to reveal a coronal source of 250--800 keV hard X-rays with a spectrum that was significantly harder than the emission from the footpoints and consistent with production by $\sim$ MeV electrons.  We extended their studies to cover a broader range of energies and showed that the spatially-integrated PLexp spectrum is in good agreement with the spectrum of these coronal $\gamma$-rays, while the sum of the spatially-integrated PL and nuclear-line spectra is consistent with the spectrum observed from the flare footpoints.  Thus, at least in this one flare where imaging data are available, the PLexp component is found to originate in the corona.  

Our study of the heliocentric-angle dependence of the PLexp index, rollover energy, and flux, Flux$_{PLexp}$, in the strong nuclear-line flares indicates that the PLexp radiation is consistent with being isotropic.  This contrasts with the anisotropic nature of the PL emission revealed by the dependence of its index and flux on the flare's heliocentric angle.  The PLexp spectrum is flat at low energies with indices that are consistent with $\lesssim$ 1. The distribution of its rollover energies peaks near 1.5 MeV, extends up to $\sim$ 5 MeV, and has a mean value of 2.2 MeV.  The $>$ 0.3 MeV flux in the PLexp component is typically less than $\sim$ 20\% of the heliocentric-angle corrected flux in the PL component. 

In a study of flares nearly 100 times weaker (see Appendices \ref{sec:weak} \& \ref{sec:eicorr}) than the 25 strong nuclear-line flares we find a similarly hard low-energy PLexp spectrum and a flux 20\% smaller relative to the PL.  Significantly, we measure a rollover energy of $\sim$900 keV that is more than a factor of two smaller than the mean rollover energy found in the large flares.  In contrast, the mean PLexp rollover energy for `electron-dominated' episodes \citep{rieg98} is more than a factor of three larger than in the large nuclear-line flares, while its flux relative to the PL flux is $\sim$ 30\% higher.

%The low-energy PLexp spectrum of flares nearly 100 times weaker (see Appendices \ref{sec:weak} \& \ref{sec:eicorr}) than the 25 strong nuclear-line flares is as hard and its flux relative to the PL flux is only 20\% smaller, but its rollover energy is $\sim$900 keV, more than a factor of two smaller than the mean energy in the large flares.  

We consider two possible origins for this hard PLexp component: Compton scattering and thin-target bremsstrahlung.  Compton scattering of a distribution of 1 keV flare X-rays by isotropic electrons with flat power-law spectra (index $\sim$ 0.1) and rollover energies between 10 and 20 MeV can produce spectra consistent with the observed PLexp component, including indices $\lesssim$ 1.  

While such a flat spectrum is a concern for a bremsstrahlung origin, we nevertheless obtained thin-target fits with comparable probabilities to those using the PLexp function.  This is primarily due to the intense PL component and its effect on the fits.  We characterized the spectrum of electrons producing the PLexp spectrum via thin-target bremsstrahlung. Our fits reveal that the electron spectrum at energies above 300 keV is flat or rises to a peak between 3 and 5 MeV and then falls at higher energies.  

Such a spectrum suggests the depletion of low-energy accelerated electrons and/or the presence of relatively persistent $\sim$ 4 MV electric fields in the corona during flares.  \citet{park97} discuss how such flat electron spectra can be produced between a few hundred keV and several MeV using three stochastic acceleration models for trapped electrons near the loop top.   Electric-field acceleration of electrons in the solar environment has been under study for decades (e.g. \citet{holm85,holm92}).  \citet{litv00} showed that MV electric field strengths can be produced in reconnecting current sheets in the corona near singular lines of magnetic field where the electric and magnetic fields are co-aligned.  Recently, reconnection electric fields of $\sim$ 4,000 V m$^{-1}$ were inferred near the X point from imaging microwave observations \citep{flei20,chen20}. Similarly-peaked electron spectra but from DC potential drops of 10 kV in the Earth's auroral zone were referred to by \citet{lin87} in their paper on solar hard X-rays bursts. 
 
 %Also close to home, $\gamma$-ray spectra from lightning-accelerated electrons, and observed by {\it RHESSI} in space as Terrestrial Gamma Ray Flashes (TGRFs), have exponential shapes with rollovers near 7 MeV and have been successfully modeled by runaway breakdown simulations \citep{dwye05}.
%Electron distributions with this low value of $s_e$, 0.1, may be produced if they are magnetically trapped, where low-energy electrons can escape more efficiently than high-energy electrons. 

The hard population of coronal electrons revealed in this study may be detectable by observations made at other wavelengths.  \citet{bai76} calculated the microwave spectrum from an $\sim$ 4-MeV exponential-shaped distribution of electrons and showed that it could produce a flux of synchrotron radiation observable at frequencies above 30 GHz. \citet{kauf04} discovered a new submillimeter component in flares with a steeply-rising spectrum at energies above 200 GHz.  Such emission can be produced by a variety of mechanisms \citep{silv07,trot08,kruc13}, including synchrotron emission \citep{bast98} and thermal bremsstrahlung from an optically thick source \citep{ohki75}.   The $\sim$ 3-15 MeV population of electrons suggested by our study have cyclotron frequencies of hundreds of GHz in coronal magnetic fields and thus could produce this submillimeter radiation via synchrotron emission.  

In Figure \ref{212_th} we compare the time histories of observed submillimeter fluxes (solid black curves) with the PL, (blue-filled circles) and PLexp (red-filled circles) fluxes in three flares where joint measurements were made.  Both the PLexp and submillimeter emissions peak later and last longer than the PL emission.   In fact, the time histories of the PLexp and submillimeter fluxes in the 2003 November 2 are remarkably similar.   Thus, it is possible that synchrotron emission from the coronal MeV electrons found in the $\gamma$-ray studies may be responsible for the rising spectrum of submillimeter emission observed in some flares.

\begin{acknowledgments}

We mourn the loss of our colleague Richard Schwartz, who led the development of the OSPEX analysis system that made our analysis of $\gamma$-ray data from {\it SMM}, {\it RHESSI}, and {\it Fermi}  possible.  Kim Tolbert played a key role in supporting our OSPEX studies and providing the Fermi solar data from the publicly available NASA archive.  J.D.F. is supported by the Office of Naval Research.  We value our discussions with Jeff Reep, Gregory Fleishman, Ivan Oparin, Albert Shih, Edward Kontar, Stephen White and the support from our colleagues at the Naval Research Laboratory.  We especially thank an anonymous referee whose comments prompted us to make the paper more readable and focussed.

\end{acknowledgments}

\appendix 

\section{{\it SMM}/GRS Instrument Response} \label{sec:resp}

 \begin{figure}[h!]
\plotone{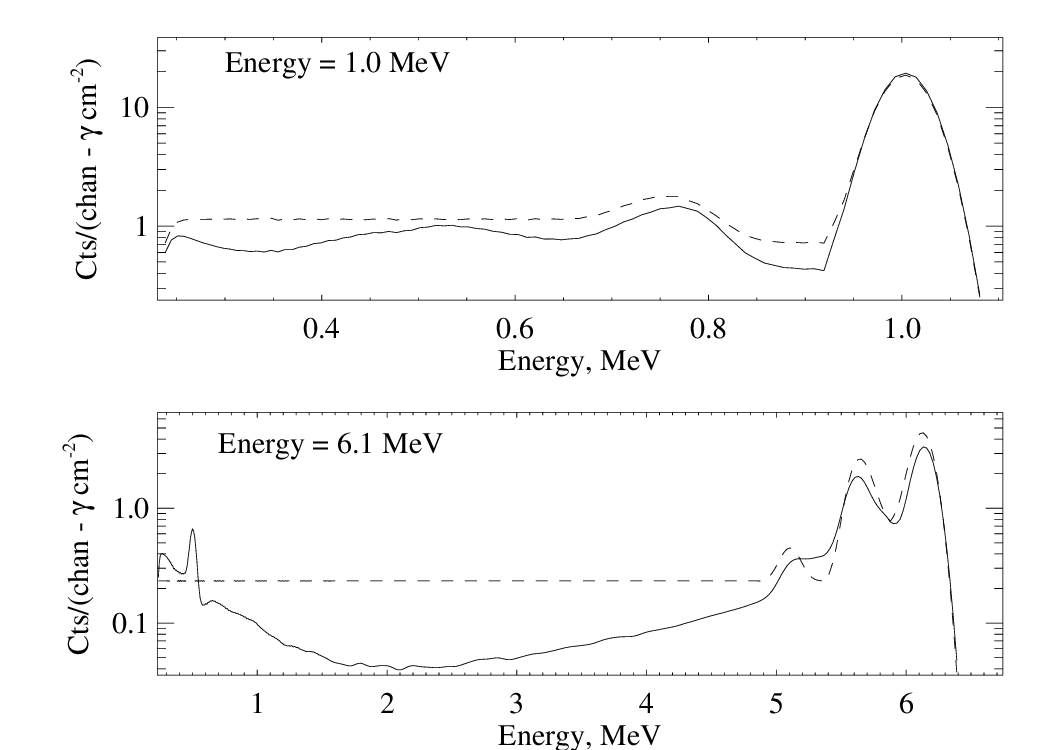}
\caption{Comparison of the original GRS empirical instrument response (dashed curve) with the Monte Carlo derived response (solid curve) for lines at 1.0 and 6.1 MeV.
\label{drm}}
\end{figure}

The detector response matrix (DRM) for {\it SMM} GRS, developed empirically before the launch of the satellite in 1980 \citep{forr80}, has been used in studies of flares (e.g. \citet{shar95,vest99}).  As part of the current investigation we reevaluated the DRM using a Monte Carlo routine that was successfully applied to the OSSE instrument on the {\it{Compton Gamma Ray Observatory (CGRO)}}) \citep{john93}.  The GRS instrument was made of seven 7.6 cm diameter x 7.6 cm long cylindrical NaI (Tl) detectors in a close-packed array enclosed by an anti-coincidence system.  The anti-coincidence system was made up of four elements: a 2.54 cm thick CsI annulus made in four sections, a 7.6 cm thick CsI plate below the NaI detectors, a plastic detector above these detectors, and one below the CsI plate.  In our simulations we assumed that the CsI detectors rejected any events with energy losses $>$100 keV.  We also simulated the effects of gamma-ray scattering from a massive aluminum spacecraft and allowed for leakage through the anti-coincidence system that would be expected from the spaces between the various elements.  

In Figure \ref{drm} we plot a comparison of the newly-adopted and original-empirical DRMs for incident photons of 1 and 6 MeV.  The most significant differences occur for photon energies above an MeV where the new photo- and escape-peak efficiencies are slightly lower and the Compton-scattered continuum is significantly reduced.  We believe that the large amount of continuum in the previous DRM, below $\sim$ 5 MeV for 6.1 MeV photons in the lower plot, originated from uncorrected room-scattered radiation from radioactive sources used in the original calibration process.  

%We compared the original and new instrument responses by fitting spectral data from three large solar flares.  In these fits we used continuum functions and nuclear templates that we discuss in the next two sections in our evaluations.  Although we found that both DRM's provided acceptable fits to the flare spectra based on the $\chi^{2}$ statistic the new response function provided significantly improvement.   Fits with the new DRM yielded values of $\chi^{2}$ that were 12, 12, and 40 below those derived using the old DRM.  If we follow the formulation of (Lampton 1976) and assume that a simple change in the detector response matrix is a equivalent to modifying the value of one relevant parameter, then we conclude that the new instrument response is an improvement over the old one at the 99.9\% confidence level. 

\section{`Electron-Dominated' Episodes}{\label{sec:rieger}

\begin{figure}[h!]
%\plotone{Thick_elec.eps}
\centering
\includegraphics[width=125mm]{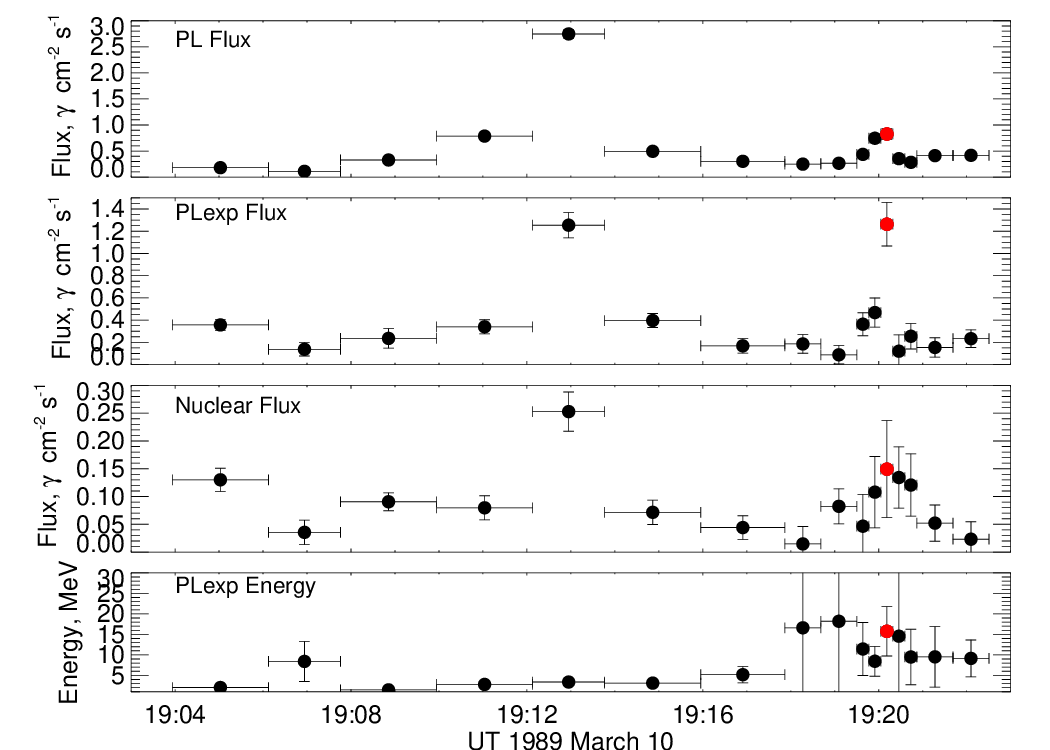} 
\caption{Time histories of the PL, PLexp, and nuclear fluxes, and the exponential rollover energy, E$_R$ determined by fits to the 1989 March 10 flare prior to and during the 8-s `electron-dominated' episode reported by \citet{rieg98}.  The red-filled circles denote the measurements made during the 16-s spectral accumulation that included the 8-s episode.
\label{89mar10th}}
\end{figure} 

One way to reveal flare electrons at MeV energies is to search for times in flares when the nuclear emission is relatively weak.  \citet{rieg98} reviewed 12 such events that they called `electron-dominated' episodes.  These events exhibit a flattening in the MeV continuum followed by rollovers at energies as high as several tens of MeV in the spectra of some time intervals.  Such features cannot be explained by transport effects alone \citep{petr94} and therefore must be intrinsic to the acceleration mechanism. \citet{park97} showed that these features can be explained using models based on stochastic acceleration by turbulence, once loss mechanisms are properly included.  Alternatively, \citet{litv00} demonstrated that electrons can be preferentially accelerated over protons to MeV energies, and above, in reconnecting current sheets.  

Here we study the spectra of these `electron-dominated' episodes to determine how the  PL, PLexp and nuclear components compare with those observed in nuclear line flares.  We begin our discussion of these events with the 8-s episode that occurred late in the 1989 March 10 flare. Notably, this flare is one of the 25 with nuclear lines in our study (Table \ref{tab:sum}).  We accumulated spectra over different durations to reveal both  the general evolution of the flare and the details of the episode.   In Figure \ref{89mar10th} we see that the PL, PLexp, and nuclear fluxes followed one another through the primary peak of the flare at 19:13 UT.  However, while the PL and nuclear fluxes rose together to an unremarkable $\sim$50-s long peak beginning at $\sim$19:19 UT, the PLexp flux increased abruptly by a factor of three in the 16-s spectral accumulation that included the 8-s `electron-dominated' episode.  During this 16-s accumulation, the PLexp component had a rollover energy, E$_R$, of $\sim$15 MeV, six times higher than the average for the entire flare (Table \ref{tab:sum}).   These observations indicate that the `electron-dominated' episode in this flare was mostly due to the abrupt increase in the flux and exponential rollover energy of the PLexp component.  There was no evidence for a decrease in the nuclear emission during this episode.

\begin{figure}[h!]
\gridline{\fig{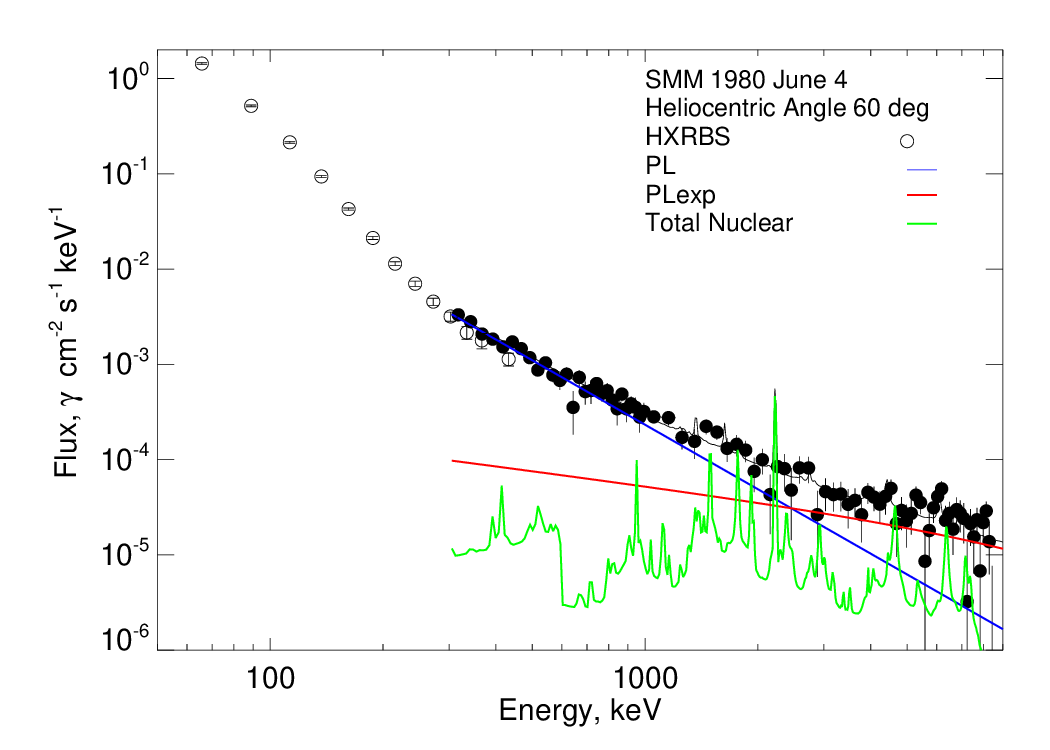}{0.5\textwidth}{(a)}
         \fig{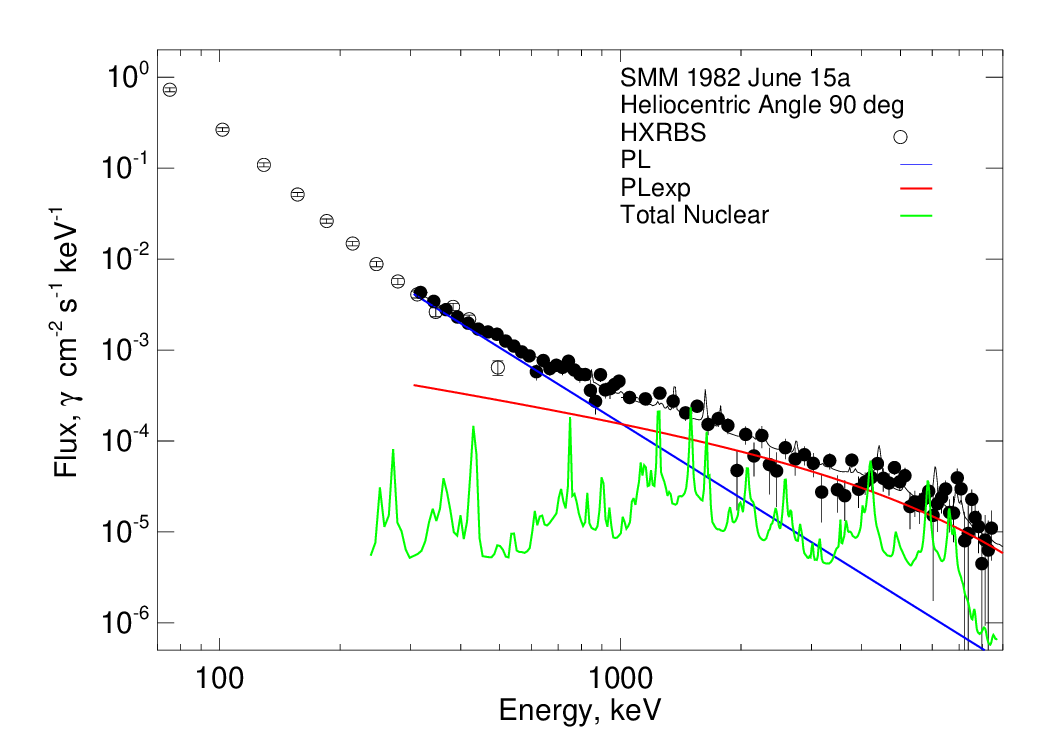}{0.5\textwidth}{(b)}}  
\caption{Combined {\it SMM} HXRBS and GRS spectra of two `electron-dominated' episodes \citep{rieg98}: (a) the 48-s long 1980 June 4 flare discussed by \citet{denn88} and (b) the 84-s long flare on 1982 June 15.   The three fitted components are plotted by the colored traces listed in the legends. 
\label{rspectra}}
\end{figure}

In Figure \ref{rspectra} we plot photon spectra from 60 keV to 8.5 MeV for two `electron-dominated' flares listed in \citet{rieg98} using HXRBS and GRS data.  As we found for the three flares plotted in Figure \ref{phspectra}, there is good agreement between the two instruments in their overlapping energy range.   The PLexp component is clearly dominant at energies $\gtrsim$ 2 MeV in both episodes with rollover energies of $\sim$ 15 and 6 MeV.  There is also a significant flux of nuclear $\gamma$ rays in both episodes.  Notably, there is something different about the hard X-ray spectra measured by HXRBS in these two flares compared with those plotted in Figure \ref{phspectra}.  In addition to the rollover in the spectra between 100 and 200 keV, both flares show evidence for hardening $\gtrsim$ 300 keV not observed in the other spectra.  From $\S$\ref{sec:plexp_par} and the above, we conclude that `electron-dominated' episodes occur when the PLexp component is more intense and its rollover energies are significantly higher than during normal flares.    

\section{Weak Solar Flares}\label{sec:weak}  
We studied the spectral characteristics of weak events by identifying 48 flares in the {\it SMM}/GRS catalog \citep{vest99} with no detectable emission above 1 MeV.  Because of the excellent gain stability of GRS \citep{forr80}, it is possible to sum up the spectra from these flares over the entire 9.5-year mission without significant degradation in spectral resolution.

\begin{figure}[h!]
\centering
\includegraphics[width=125mm]{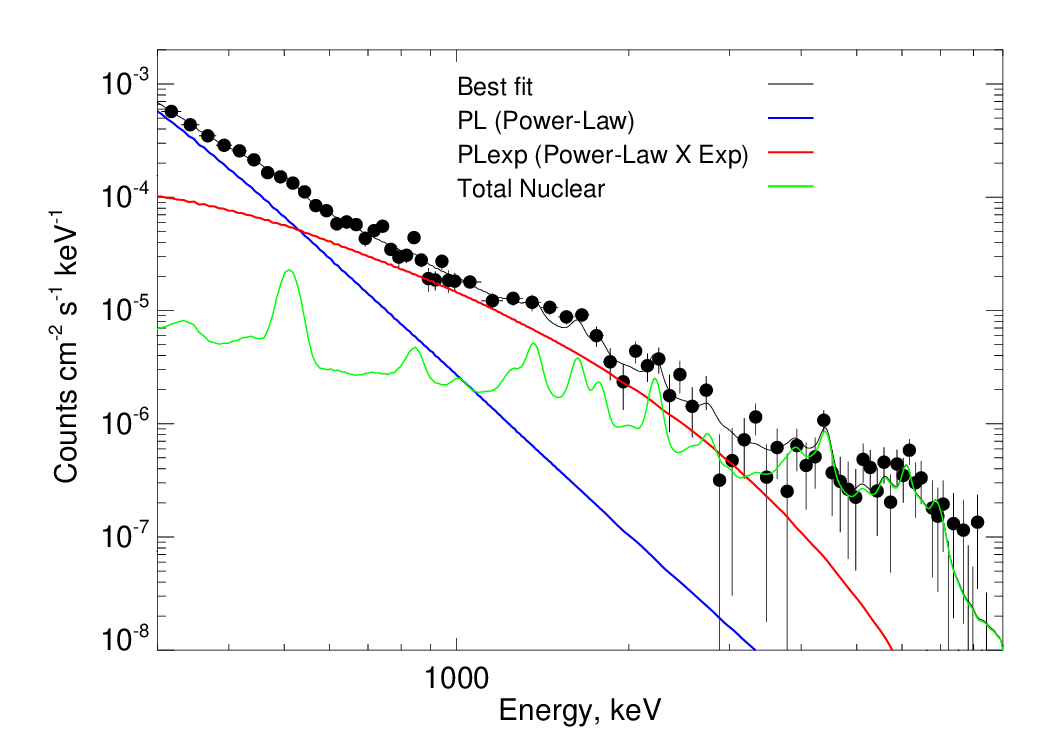} 
\caption{Summed spectrum from 48 flares observed by {\it SMM}/GRS each having no detectable emission $>$1 MeV.  The fitted PL, PLexp, and nuclear components are plotted and defined in the legend.  We used broader energy bins at high energies to improve statistics.
\label{33flrsum}}
\end{figure}

In Figure \ref{33flrsum} we plot the summed count spectrum of these 48 flares.  With the improved statistics at high energy from this summation we see that the spectrum of these weak flares clearly extends up to the 8.5 MeV maximum energy measured by GRS.  \citet{shar00} presented a similar spectrum from weak flares but did not have the improved instrument response and nuclear templates to properly fit the data. Our fit to the summed spectrum reveals a PL component (blue line) that dominates below 500 keV and has an index of $\sim$ 3.5, the same as the mean index of the 25 large flares studied in this paper.  The fit also reveals clear evidence for both the PLexp (red curve) and nuclear-line (green curve) components found in the more intense flares.   What we find surprising is the absence of a significant 2.223-MeV neutron-capture line. The 2.223-MeV/nuclear de-excitation line flux ratio is 0.01 $\pm$ 0.006 in the sum of 48 weak flares.  This is a factor of 40 smaller than the ratio observed in the 25 nuclear-line flares.  Because the cross section for producing the neutron-capture line peaks at a higher-energy than the cross-sections for producing nuclear de-excitation lines, such a reduction in 2.223-MeV flux suggests that the ion spectrum in weak flares is significantly steeper than in larger flares.  We note that the lack of significant 2.223-MeV line emission in our summation of weak flares may be inconsistent with the good correlation between the line and bremsstrahlung fluences found by \citet{shih09} (see Appendix \ref{sec:eicorr}).

\section{Power-Law Electron-Produced \textit{vs} Ion-Produced $\gamma$-Ray Fluence Correlations}\label{sec:eicorr}

\begin{figure}[h!]
\includegraphics[width=125mm]{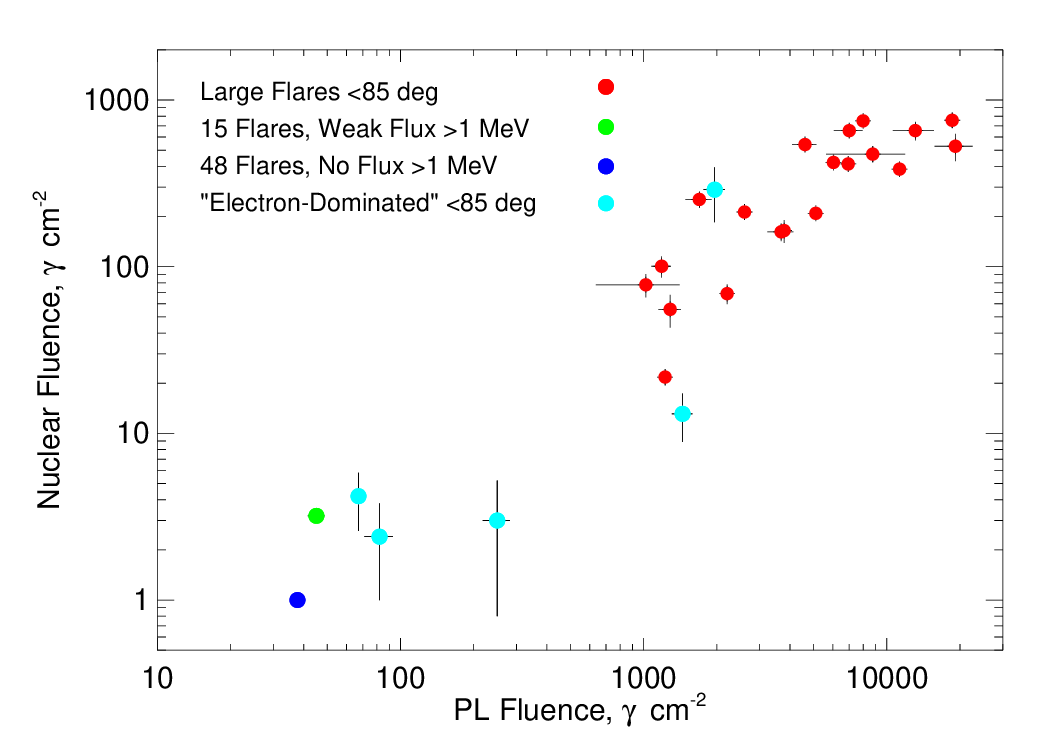} 
\caption{Plot of the nuclear de-excitation line \textit{vs} the angle-corrected PL fluences for flares.   Plotted are the fluences from 20 nuclear-line flares (red), 5 `electron-dominated' episodes, the average fluence of the sum of 15 flares (green), each with only weak flux detected $>$1 MeV, and the average fluence of the sum of 48 flares (blue), each with no detectable flux $>$1 MeV. 
\label{correlations}}
\end{figure} 

\citet{shih09} analyzed a large sample of {\it RHESSI} and {\it SMM} flares and found a close correlation between the $>$300 keV electron bremsstrahlung fluence and 2.223 MeV neutron-capture line fluence from ion reactions over three orders of magnitude.  Earlier studies revealed the close correlation between electron bremsstrahlung and nuclear de-excitation line fluences \citep{vest88,murp93}.  These correlations indicate that accelerated electrons with energies $>$300 keV and protons with energies $\gtrapprox$2 MeV have a related origin.  We note that in these earlier studies the fitted bremsstrahlung fluences were contaminated with a residual nuclear line and continuum, because their fits used Gaussian line approximations and not the latest theoretical nuclear spectra \citep{murp09}.  In addition there was an instrumental nuclear-line component that was not corrected for in the {\it RHESSI} response matrix (see discussion in $\S$\ref{sec:fits}).  Note also that the bremsstrahlung fluence in the earlier studies was not corrected for the heliocentric angle of the flare (see $\S$\ref{subsec:iso}).

We repeated these earlier comparisons using the angle-corrected PL and the nuclear de-excitation line fluences in 20 of the large flares at heliocentric angles $<$85$^{\circ}$. We restricted the sample to avoid bremsstrahlung attenuation effects at the limb.  The results of this comparison are plotted as red-filled circles in Figure \ref{correlations}.  We extended the plot to much weaker fluences using the summed spectra of the 48 weak flares (blue-filled circles) discussed in $\S$\ref{sec:weak} and 15 flares that were individually detected with weak emission $>$1 MeV (green-filled circles).  The fluences of the `electron dominated' episodes fill in the intermediate region.  With these measurements, we see that the nuclear-line and PL fluences are correlated over almost three orders of magnitude.

\bibliography{references}{}
\bibliographystyle{aasjournal}

\end{document}